# A review of plasma–liquid interactions for nanomaterial synthesis

Qiang Chen,[1] Junshuai Li[2] and Yongfeng Li[3]

[1]Fujian Provincial Key Laboratory of Plasma and Magnetic Resonance, Institute of Electromagnetics and Acoustics, Department of Electronic Science, Xiamen University, Xiamen 361005 China

[2]School of Physical Science and Technology, Lanzhou University, Lanzhou 730000, China

[3]State Key Laboratory of Heavy Oil Processing, China University of Petroleum, Changping, Beijing 102249, China

Email: chenqiang@xmu.edu.cn

**ABSTRACT:** Over the past decades, a new branch of plasma research, the nanomaterial (NM) synthesis by plasma-liquid interactions (PLIs), is rapidly rising, mainly due to the recently developed various plasma sources operated from low to atmospheric pressures. The PLIs provide novel plasma-liquid interfaces where many physical and chemical processes take place. By exploiting these physical and chemical processes, various NMs ranging from noble metal nanoparticles to graphene nanosheets can be easily synthesized. The currently rapid development and increasingly wide utilization of the PLI method naturally lead to an urgent requirement for presenting a general review. This paper reviews the current status of research on plasma-liquid interactions for nanomaterial syntheses. Focus is on the comprehensive understanding of the synthesis process and perceptive opinions on the present issues and future challenges in this subject.

## 1. Introduction

Plasma is one of the four fundamental states of matter besides solid, liquid, and gas, and they are closely relevant to the human life and modern industry. Plasmas resulting from ionization of neutral gases, generally contain an equal number of positive ions and negative electrons (negative ions in some cases), in addition to neutrals, metastables, excited atoms or molecules, reactive radicals, ultraviolet light, and strong electric field. According to the flexible reactivity of the species in plasmas, the gas-based reactive plasmas are widely used in manufacturing industries such as surface modification of materials, surface processes in the integrated circuits processing [1].

Over the past decades, a new branch of plasma research, the nanomaterial (NM) synthesis by plasma-liquid interactions (PLIs), is rapidly rising, mainly due to the recently developed various plasma sources operated from low to atmospheric pressures. In the PLIs, plasmas are over or inside liquids, providing plasma-liquid interfaces where many physical and chemical processes can take place, and these processes can be used to synthesize various NMs. In fact, the idea of forming materials from the PLIs is not new, it appeared long before the term "plasma" was used to describe a glowing discharge by Irving Langmuir in 1928 [2], and they can be dated back to 1887 when Gubkin [3] used a discharge to reduce silver ions ($Ag^{+}$) in an aqueous solution of $AgNO_3$. After this inspiring work, only a few studies were followed for material synthesis from the PLIs in the early time, such as iodine production from aqueous solution of potassium iodide [4], formation of hydrogen peroxide from dilute sulphuric acid solutions [5, 6], and amino acid synthesis from elemental carbon related materials [7-10]. The reason might be no interest for researchers to form materials and lack in requisite equipment to probe the formed materials (usually in a scale of nanometers) at that period. On the other hand, much interest was paid for the electrolysis induced by the PLIs, which has been comprehensively explored by Hickling et al. [6, 11-13] using the liquid as cathode. The violation of Faraday's law about the charge transfer in a usual electrolysis was observed, i.e., the chemical effects produced by this system is far larger than that predicted by Faraday's law [5, 14], which is explained by the fact that there exists complicated chemical reactions induced from the liquid irradiation by energetic ions (in plasma) at the plasma-liquid interface besides charge transfer. In addition, very recently, a well-documented review was presented to summarize the charge transfer processes at the plasma-liquid interface [15].

All the pioneering works demonstrate that a great number of physical and chemical processes exist in the plasma-liquid system involving the bulk plasma, the liquid as well as the plasma-liquid interface. Therefore, many applications can take advantage of the unique properties of the PLIs including drinking and waste water treatments [16-19], hydrogen peroxide production [11, 20], plasma medicine [21-24], and liquid analysis [25-34] etc. It is well known that NMs are bridges between bulk materials and atomic or molecular structures, making them of great scientific interest. Different from their bulk counterparts, for example, the properties of nanoparticles (NPs) are usually size-dependent, and when the particle's size approaches nanoscale, the number ratio of surface to bulk atoms increases rapidly, which usually affect the quantum confinement in semiconductor particles [35], localized surface plasmon resonance (LSPR) in metal particles [36], superparamagnetism in magnetic materials [37], the melting temperature [38] as well as the catalysis capability [39] of the materials. A successful route to synthesize NMs, solution-based synthesis, is based on etching and/or reduction of metal ions in solutions [40, 41], which has proved a powerful tool for forming NMs with various sizes, shapes as well as compositions. Since the plasma can provide electrons and radicals, and the metal ions can dissolve in the liquids, this unique plasma-liquid system is strongly attracted to the NM syntheses, and they fertilize and active a rapidly rising field of the PLI-based NM syntheses [42-49] in recent decades due to the strong requirement of large-scale syntheses of NMs and the fast development of various plasma sources.

In a vacuum system, dendrites of several metals have been obtained in 1970 by a glow discharge electrolysis of molten metal salts [50]. $C_{60}$ has been formed by electric discharges in liquid toluene [51] in 1993. In 1998, Sano et al. [52, 53] reported a high-rate carbon "onions" fabrication by using graphite arc discharge in water. To our knowledge, the very first demonstration for the synthesis of Ag fine particles was performed by Kawamura et al. [54] in 1998, and they have shown the synthesis of microscale fine Ag particles by operating an atmospheric-pressure Ar direct-current (DC) plasma over a molten LiCI-KCI-AgCI eutectic, where the molten salt acted as anode. The $Ag^{+}$ in the molten salt was reduced to zero-valent Ag by reducing species in the plasma and then fine Ag particles were formed by the nucleation and growth of the zero-valent Ag in the molten salt. In principle, the electrons generated from the plasma cathode can reduce any metal ions in its molten form. This prototype, plasma as cathode and liquid as anode, is later taken in many cases for synthesizing NMs.

Reducing agents can be automatically produced during the NM synthesis via the PLIs, which is a key advantage in contrast to conventional, solution based methods. The plasma includes not only free electrons and ions with certain energies dependent on the discharge mode, but also radicals produced by plasma-liquid and/or gas interactions. Besides the free electrons, the ions and radicals can cause numerous physical and chemical reactions in the plasma-liquid interface and in the liquid. The NM formation can be attributed to the complicated physical and chemical processes in the PLIs, i.e., the reduction, oxidation, and sputtering etc. The synthesis process can be controlled not only from the solution phase as in a usual solution-based synthesis, but from the plasma phase, i.e., adjusting the reducing species yields by tuning the plasma parameters. Because the reducing species produced from the PLIs have different lifetimes and different reducing abilities [55], the process of NM synthesis can also be tuned by tailoring the relative yields of specific reducing species from the plasma part. Although the diversity of reducing species in the PLIs increases the complexity for analyzing the synthesis mechanism, on the other hand it expands the routes to tune the synthesis process. Moreover, the versatile radicals in the PLIs render a possibility to control the shape and composition of NMs by adjusting plasma parameters.

At present, there are well documented reviews on the other applications of the PLIs [18-24] rather than the NM syntheses. Although a few review papers on the NM syntheses by the PLIs [56-61] can be found in the literature, they are either limited to the authors' own works or less in comprehensive understanding and generality. We consider that we have passed the rudimentary period of the NM syntheses by the PLIs, there is an urgent requirement to present a general review involving a comprehensive understanding of the synthesis process and perceptive opinions on the promising future.

This review aims to present a comprehensive understanding of the physical and chemical processes taking place during the NM synthesis by the PLIs, and summarizes the recent improvements and progresses in this field. We organize the content as follows. After this general introduction, Section 2 presents a brief introduction to the physical and chemical processes which occur in the PLIs. Section 3 summarizes the achievements of the NM synthesis from the PLIs. Section 4 points out the present issues need to be addressed in the NM synthesis by the PLIs and discusses the possible solutions. Finally, we summarize our points and give an outlook for the future work.

## 2. Physical and chemical Processes in Plasma-Liquid Interactions

Actually, the chemical reactions in NM syntheses from the PLIs are fairly simple, most of the cases are based on the reduction of metal or semiconductor ions by electrons provided from the plasma species, except some cases the building blocks of NMs originate from simply direct physical processes such as sputtering or evaporation of target materials without chemical processes. Coupling the complexness of both the plasma and liquid, the plasma species able to provide reducing electrons are quite intricately dependent on the gas type, the type of power source, the discharge pressure, liquid properties, and the electrodes' configuration, etc.

The physical and chemical processes in pure gas-phase plasma have been intensively studied by many researchers and technicians in order to elaborately control the processes for numerous industrial applications. But the studies on physical and chemical processes in the PLIs have attracted researchers' attention just in recent decades due to the beneficial applications in water treatment [16-19], plasma medicine [21-24], the NM synthesis [42-49], surface engineering [62-65] etc.

Although there are various types of plasma-liquid systems, they can be simply classified as plasma over the liquid and plasma inside the liquid. In both classes, there is a plasma-liquid interface in addition to the liquid phase and the gas-phase plasma. The plasma-liquid interface is an important zone for NM syntheses at which many physical and chemical reactions can be induced by the PLIs. In this section, we will give a comprehensive analysis of the physical and chemical processes which appear in the PLIs. The synthesis mainly based on the physical processes is discussed after the synthesis mainly based on the chemical processes. It is well known that the power sources to drive a plasma come in many forms, including DC, pulse, microwave, radio frequency etc., but here we focus our discussion on the plasmas driven by the DC source. These conclusions can be partially extended to other sources. In the following part, we seek to give a clear image by illustrating the possible processes.

## 2.1. Chemical Processes

### *2.1.1. Plasma over Liquid*

In this case, the analysis will be carried out by arranging DC plasma with liquid acting as anode or cathode. The results can be generalized to many similar cases.

Kaneko et al. [66, 67] investigated the plasma potential distributions for a DC plasma-liquid system at low pressure in Ar atmosphere. The high vapor pressure liquids, such as water or ethanol, are unstable at low pressures due to evaporation. Therefore, volatile liquids are not suitable for the low-pressure plasma-liquid research. Thanks to the negligible vapor pressure of ionic liquids (ILs) [68], the authors used ILs in the experiments. The potentials in plasma and in ionic liquid are measured by a Langmuir probe and an electrostatic probe, respectively. Figure 1 presents the plasma potential results for both the IL as cathode and anode. Similar to a traditional solid-electrodes system, the authors found that there is a large cathode fall at the IL surface when the IL acts as cathode, while there is no voltage fall at the IL surface as the IL is anode. The results were confirmed by optical emission spectra detected from plasma, implying an irradiation of energetic ions onto the IL surface at the liquid-cathode case, and low energy electrons shower onto the IL surface at the liquid-anode case. These results are very important on understanding the processes in the PLIs. Because many experiments of the plasma-liquid system use aqueous solutions, we present the detail processes by exemplifying a plasma-aqueous solution system as follows.

There are many partial interpretations [42, 44, 46, 48, 55, 57, 69-73] on the NM syntheses from the PLIs, here we summarize them and try to give a general mechanism. It is worth pointing out that the plasmas are operated at tenths kPa to atmospheric pressure and they usually have a small size and low gas temperature due to the volatility of water. As shown in Fig. 2, Ar plasma is generated over an aqueous solution of a metal salt, an inert metal plate such as Pt is immersed in the solution to conduct the circuit. The physical and chemical processes are illustrated for the plasma-liquid system with the aqueous solution as cathode [Fig. 2(a)] and anode [Fig. 2(b)]. In the solution, there are dissolved metal ions ($M^{n+}$) which are the precursors of the final MNPs. The chemistry can be simply expressed as

$$\mathrm{M}^{n+} + ne^{-} = M. \quad (1)$$

However, the electrons provided by the plasma can come from very different ways.

Figure 2(a) presents the processes for the liquid cathode case, as indicated in Fig. 1(a), there is a large cathode voltage fall at the liquid surface, the Ar ions in the bulk plasma will be driven by this voltage fall to move toward the liquid surface, when they reach the surface, they will hold an energy in the order of tens to hundreds eV, depending on the gas pressure and the discharge voltage. Then the energetic ions attack the liquid surface, creating some secondary effects, such as generating secondary electrons, decomposition of the constituents in the liquid. In addition to sustain the bulk plasma, the secondary electrons can dissolve into the water to form hydrated electrons ($e_{aq}^{-}$) which have very strong reducing ability [74]. In the aqueous solution case, water molecule will be decomposed to atomic H and OH radical which have been confirmed by the optical emission spectra detected near the liquid surface [55]. Atomic H can combine with each other to form $H_2$. Besides direct combination of OH radicals, $H_2O_2$ can be formed in the PLIs by many different pathways [20]. Atomic H and $H_2$ are able to reduce metal ions in the solution. In addition, depending on the pH values, $H_2O_2$ presents different reducing abilities [75-77] for certain metal ions although it is usually a strong oxidizing species, for example, gold (III) reduction by $H_2O_2$

$$3H_2O_2 + 2Au(III) \rightarrow 2Au(0) + 3O_2 + 6H^+ \quad (2)$$

Dissociative electron attachment can lead to the formation of hydride ($H^-$) [78] which is a very strong reducing species and can be dissolved into the liquid.

$$H_2O + e^- \rightarrow H^- + OH \quad (3)$$

Ultraviolet (UV) light is usually observed in the plasma-liquid system [79, 80]. Therefore, the high energy of UV photons can be transferred to water molecules, producing potential reducing/oxidizing species,

$$H_2O + h\nu \rightarrow H + OH \quad (4)$$

The solution components can be forced into the plasma-liquid interface even into the bulk plasma by sputtering and/or evaporation [81-83], and this fact is usually ignored by researchers when considering the metal ion reduction by the PLIs. The solution components will take a form of droplet in the plasma-liquid interface and/or in the bulk plasma due to the transport by sputtering and/or evaporation. The droplets have large ratio of surface to volume, and therefore they can meet much more electrons (including the bulk-plasma elections and electrons produced from the reactive radicals) compared with the liquid at the surface, leading to a more efficient reduction of metal ions in the droplets.

For the liquid-anode case, as indicated in Fig. 1(b), there is no voltage fall at the liquid surface, just electrons from bulk plasma shower onto the liquid, and the electrons can dissolve into the liquid to form $e_{aq}^-$. In addition, atomic H, $H^-$, OH, $H_2$, and $H_2O_2$ can be produced in the bulk plasma by decomposing water vapor and cascade reactions, but the yields should be much less than those in the liquid-cathode case. UV light can also function as the energy source for producing atomic H and OH. Due to the lack of strong irradiation by energetic ions on the liquid surface, fewer droplets can be formed in the plasma-liquid interface and the bulk plasma, and the metal ions in the droplets are possible to be reduced by electron providers around them.

As discussed above, the yield of reducing species generated the liquid cathode is much larger than that in the liquid anode, although they can generate the same reducing species. Thus, the reducing efficiency in the liquid cathode is higher than that in the liquid anode. This phenomenon can somehow influence the size and structure of the final products [40, 84, 85]. In addition to the reducing species, the oxidizing species, atomic O [86], $O_3$ [87], OH radical, etc. can also be generated in the PLIs. All species can diffuse into the liquid. OH radicals are easily combined into $H_2O_2$, then pass through the liquid surface and drift in solution. $H_2O_2$ as well as dissolved $O_2$ can act as oxidizing species. As a result, reactive metals such as iron, cobalt etc. might be oxidized after the formation of metal NPs by reduction of metal ions. In order to give a clear description, we summarize the standard reduction potential ($E^o$) vs standard hydrogen electrode (SHE) values of partial reactive species in the PLIs in Table 1 in which some species we do not mention in the previous analysis are also included.

Free electrons from bulk plasma, secondary electrons from ion irradiation on the liquid surface have much strong reducing abilities due to the their high energies, they can reduce almost any metal ion in their molten forms or most of metal ions in aqueous solution. For other reducing species, we can estimate their reducing ability by their $E^o$. The $e_{aq}^-$ from electrons' solvation are very strong in reducing ability ($E^o$=-2.87 V vs SHE [88]) although it has a short lifetime (~ns to

~μs depending on the medium [89]). Based on the $E^o$ data in Ref. [90], we can roughly estimate that the $e_{aq}^-$ can spontaneously reduce most metal ions even the much weak oxidants of $Fe^{2+}$ and $Mg^{2+}$ in an aqueous solution, while the spontaneous reduction of $Cs^+$ ($E^o$=-3.03 V) by $e_{aq}^-$ is not feasible. If the solution interacted with plasma is composed of chloroauric acid ($HAuCl_4$), then metal ions will take on a form of $[AuCl_x(OH)_{4-x}]^-$ (x≥2 at low pH values, x<2 at high pH values) [91, 92]. Considering Table 1 and taking $AuCl_4^-$ as an example, the species generated in PLIs such as OH, $H_2O$, and $H_2O_2$ can have the following possible reactions with $AuCl_4^-$ in acidic (Eqs.5.1-3) and basic (Eq. 5.4) media,

$$AuCl_4^- + \frac{3}{2}H_2O_2 \rightarrow Au + \frac{3}{2}O_2 + 3H^+ + 4Cl^- \quad (5.1)$$

$$AuCl_4^- + 3H_2O + \frac{3}{2}OH \rightarrow Au + \frac{3}{2}H_2O_2 + 3H^+ + 4Cl^- \quad (5.2)$$

$$AuCl_4^- + \frac{3}{2}H_2O \rightarrow Au + 3H_2O_2 + 3H^+ + 4Cl^- \quad (5.3)$$

$$AuCl_4^- + \frac{3}{2}H_2O_2 + 3OH^- \rightarrow Au + 3H_2O + \frac{3}{2}O_2 + 4Cl^- \quad (5.4)$$

The standard Gibbs free energy ($\Delta G^o$) [90] of a chemical reaction can be expressed as $\Delta G^o$=-$nFE^o$, where $n$ is the electron number involved in the chemical reaction, and $F$ is Faraday constant ($9.65 \times 10^4$ C mol$^{-1}$). A chemical reaction is favored if $\Delta G^o$ is negative, while it is not when $\Delta G^o$ is positive. Based on $E^o$ in Table 1, $\Delta G^o$ can be estimated to be -0.3×3 $F$, -0.04×3 $F$, +0.76×3 $F$, and -1.15×3 $F$ for Eqs. 5.1-4, respectively. Therefore, Eqs. 5.1, 5.2 and 5.4 can go forward spontaneously because of the negative $\Delta G^o$, while Eq. 5.3 cannot due to its positive value of $\Delta G^o$. From the $\Delta G^o$ values, we can also know that Eq. 5.4 is much easier to proceed than Eq. 5.1 and Eq. 5.2, which means that $H_2O_2$ is a stronger reducer for $AuCl_4^-$ in basic than in acidic medium.

The behaviors of other metal ions and reducing species can also be analyzed in the same way. Because the reducing species in plasma over liquid can just penetrate a certain depth of the bulk liquid due to their limited energy and lifetime, the chemical reactions take place mainly at the plasma-liquid interface and a limited depth under the liquid surface. We know that metal ions can be driven to move by an electric field applied in the liquid, for example, $Ag^+$ and $AuCl_4^-$ will move toward the liquid surface and bottom, respectively, in a liquid anode case, which will influences the reaction rates and final products. Thus, according to metal ions movement by the applied electric field and the reducing abilities of existing species, the selection of suitable forms of metal ions is important for the NM syntheses from a plasma over liquid system.

It is worth pointing out that the reducing species can be classified as two types in terms of their lifetimes: the short-lived and long-lived species [55]. The short-lived reducing species include free electrons, $e_{aq}^-$, secondary electrons, atomic H, $H^-$, and UV light which disappear or quickly decay after stopping the plasma irradiation. The long-lived reducing species consist of $H_2$ and $H_2O_2$ which can stay in the liquid for a long time, even able to be detected several months later after the plasma irradiation. These two types of reducing species can be used to control the nucleation and growth processes of the synthesized NMs. The pH value of solution can be also tuned to control the reaction processes in the solution since the reducing properties of $H_2O_2$ for numerous compounds are stronger in basic media than in acidic ones [75-77]. In Section 4.1, we will return to this subject for a detailed discussion.

In addition, the PLIs are usually operated in ambient air, $N_2$ and $O_2$ invariably enter the reaction zone, resulting in solution acidification through various cascade reactions of plasma induced nitrogen species in solution [93], such as the dissolution of nitrogen related species ($NO_2$, $HNO_3$ etc.) [79, 94, 95]. The $H^+$ can also be created from water ionization by electron collision. Thereby, the liquid chemistry such as pH value will be changed by the plasma irradiation, which can influence the rate and probability of chemical reactions in liquid, that we must pay attention if we want to elaborately control the synthesis. Interestingly, hydrazine ($N_2H_4$) has been observed after an exposure of $N_2$ and $CO_2$ [96] or Ar/ammonia [97] plasma on water, and Pd [98] or Cu [99] NPs have been obtained from reductions of $K_2[PdCl_4]$ or $CuCl_2$ by hydrazine. Consequently, the reducing species in PLIs operated in air might include hydrazine since both $N_2$ and $CO_2$ are components of air. But we should point out that one must be very careful if one wants to use hydrazine (hydrate) and hydrogen peroxide as a reducing agent, since the mixture of hydrazine (hydrate) and hydrogen peroxide is liable to be explosive.

Anodic dissolution has been used for electrochemical synthesis of Pd [100] and Ag [101] NPs, and gold nanorods [102], anodic dissolution can also take place in the PLIs since the plasma-liquid system is actually a modified electrochemical cell [15, 103]. While just a few studies considered the effect of anodic dissolution in the NP synthesis from the PLIs [46, 69]. We should point out that the anodic dissolution should not be omitted in controlling the NM synthesis. If the material of immersed plate electrode is the same to the dissolved metal ions ($M^{n+}$), the anodic dissolution can take place at the electrode, liberating $M^{n+}$ into the solution, which will increase the concentration of $M^{n+}$. The precursor concentration can strongly affect the size [84, 85] and structure [40] of the final products, and therefore, this anodic dissolution can add uncertainty to control the final products. On the other hand, the anodic dissolution will add impurity in the final products if the material of immersed electrode is different from the dissolved metal ions. In order to control the synthesis process, we must eliminate or minimize the anodic dissolution, which can be realized by carefully selecting the solution composition, metal type, ambient gas, and discharge type [104].

*2.1.2. Plasma in Liquid*

Different from plasma over liquid, most of the plasmas in liquid are operated in liquid bubbles or small vapor channels. Figure 3 shows two types of plasmas: plasma in bubble, and streamer plasma in small vapor channel. The discharge process of plasma in bubbles is similar to that of glow discharge plasma in a gas phase and plasma over liquid since the bubbles consist of water vapor created by joule heating and electrolysis of water. The bubble plasma usually propagates along the plasma-liquid surface or cross the bubble without contacting the liquid surface depending on dielectric constants of the bubble gas and liquid [105, 106]. While the streamer plasma in water is generated by decreasing of liquid density around the electrode by electric field enhancement [107] under a high overvoltage, or direct electron collision induced liquid water ionization under ultrahigh local electric field [108-111]. The streamer plasma takes on a form of multi-branched channels as shown in Fig. 3(c) and propagates in these small channels with very high speeds.

Although the discharge processes are different for plasma in vapor bubble and streamer in small vapor channel, they show similar reactive species in the bulk plasmas as presented in Figs. 3(b) and (d). The dominant species originate from the decomposition of water by plasma, atomic O, atomic H, OH radicals, also sometimes nitrogen related species due to residual dissolution of $N_2$ in the studied liquid, while in most cases, nitrogen related species are neglectable.

As presented in Fig. 4(a), plasma in vapor bubble will generate similar species as in a plasma over liquid (except the species derived from discharge gas, for example, N-containing species provided by discharges in air). Under some conditions, plasma localizes around the anode [112] or the cathode [113] bubble which is similar to a liquid cathode or anode case, and therefore it can be analyzed in the same manner as the plasma over liquid case. However, the reactive species generated in bubble can diffuse across the plasma-liquid interface and will enter the liquid when the bubble is broken [114], which frequently happens during the operation, and as a result, the reducing efficiency should be relatively higher compared with the plasma over liquid.

For the streamer plasmas in small channels, as described in Fig. 4(b), they can produce shock waves and cavitation bubbles [115] in addition to the same reactive species as produced by plasma in the bubble. Shock wave can transfer energy to the liquid and produce cavitation bubbles in liquid. Theses cavitation bubbles can experience a broken process and generate secondary small plasmas [116], which will enhance the yield of reducing species in the liquid.

Because both electrodes are immersed in liquid, the final products might be contaminated by the electrodes' material. Therefore, the electrode materials and solution constituents must be carefully selected in order to decrease potential impurity as less as possible.

In addition to the aforementioned reduction of metal ions, the solvent decomposition is also used for the NM syntheses from the PLIs. Since plasma can generate rich reactive species, NMs might be formed by decomposing the solvents, such as carbon sphere formation from ethanol [117] and benzene [118] decompositions, and silicon nanocrystals surface engineering [65, 119] by functional groups produced from ethanol decomposition.

It is worth noting that in both plasma over and inside liquid, reactions in nonequilibrium can take place by thermal or electron impact activation.

### 2.2. Physical Processes

The above discussions mainly concentrate on the chemical processes used to synthesize NMs from the PLIs, while there are some cases, the physical processes are dominant. Two physical processes frequently used for NM syntheses in the plasma-liquid system are presented as follows.

#### *2.2.1. Sputtering* [120, 121]

As a representative example, Fig. 5 illustrates the AuAg NPs synthesis by sputtering in a plasma-IL system. An AuAg foil target is placed over the IL, and a negative high voltage is applied to the target to generate Ar plasma between IL and AuAg target. As shown in Fig. 1(b), a high cathode voltage fall will be formed on the target after plasma generation, and this cathode voltage fall will cause target sputtering by energetic ions ($Ar^+$). Similar to a traditional sputtering deposition, the sputtered target atoms will move toward the IL surface, while theses atoms will experience nucleation and growth to NPs on the liquid surface and inside the liquid rather than form a film on a solid substrate as in a traditional sputtering deposition due to the flexibility of liquid. The final AuAg composition and size are possible to be controlled by the target composition and plasma parameters. The selection of IL is based on its low vapor pressure which facilitates the high efficient sputtering and minimize the contamination in low-pressure plasma.

*2.2.2. Evaporation* [122, 123]

Besides the sputtering process, evaporation of metal electrode is also used in the synthesis of NMs as shown in Fig. 6. In this case, a thin metal wire is partly immersed in an electrolyte solution as cathode, while another inert metal such as Pt mesh also immersed in the solution is used as anode. When a gradually increasing voltage is applied to the cathode, joule heating generates water vapor which can cover the immersed part of the thin wire. Short-lived cathode spot can be ignited at the thin wire as voltage increases to the breakdown voltage of the water vapor. As a result, there exist cathode spots on the thin wire surface. The localized temperature at the cathode spot is very high (several to tens of thousands $^{o}C$) which can result in a fast evaporation of cathode material, leaving craters on the cathode surface, which was confirmed by checking the morphology of the thin wire surface after plasma operation [124, 125]. Consequently, the evaporated cathode materials will move to the liquid surface and they will be cooled down during the movement and finally quenched in the liquid, leading to the final formation of NPs. The quality and size of NPs depend on the applied voltage as well as the electrolyte type and concentration.

## 3. Nanomaterials Synthesized by Plasma-Liquid Interactions

Refering to Table 1 and Ref. [77], we find that the maxium $E^o$ of the reducing species produced by the PLIs are larger than those of most metals and semiconductors in the perodic table, thus in principle, NMs of most metals and semiconductors can be synthesized from the PLIs. Thanks to the efforts of many reseachers' groups, a great number of different NMs have been achieved in the last two decades. We review these recent develoments and progress as follows in terms of types of materials, and each type of material is related to different configurations of plasma-liquid system.

### 3.1. Noble Metals

Au, Ag, Pt, and Pd are usually categorized as noble metals, and except Ag, they are resistant to corrosion and oxidation in moist environment. They have been used in the human life as jewelry and currency since the ancient times. When their sizes approach less than 100 nm, they display exotic optical, electronic, etc. properties, rendering numerous fundamental studies and engineering applications. Au and Ag NPs have extraordinary optical properties, the well-known local surface plasmon resonance (LSPR) which is a charge oscillation of the collective electrons at the NP's surface excited by light. The LSPR can be tailored by particle size, composition and shape, rendering Au and Ag NPs a great number of potential applications, such as surface-enhanced Raman scattering (SERS) [126], nanomedicine [127], biological sensing [39]. Although Au is chemically stable, AuNPs were found to have fascinating catalytic ability for many oxidation reactions [128, 129] when AuNPs are supported by $Co_3O_4$, $Fe_2O_3$, or $TiO_2$, and the catalytic ability is closely correlated with the size of AuNPs. Pt and Pd NPs also possess unique catalytic ability for CO oxidation [130, 131], hydrogenation [132, 133] etc. The NPs of these noble metals will play pivotal role in the development of new technologies. Therefore, synthesis of these noble metal NPs with low cost, high quality, and scale yield is necessary for scientific research and practical application. The PLI method provides a simple route to realize these aims.

In fact, these noble metals are dominant in metal NMs syntheses by the PLIs because of their strong resistant to corrosion and oxidation, and their higher electrochemical series (i.e., ions of noble metal are easier to capture electrons). Some representative examples will be presented as follows.

The synthesis of Ag nanowires has been reported in 1999 from an arc discharge generated over an aqueous solution of $NaNO_3$ with two silver filament electrodes, one over and the other immersed in the solution [134]. In 2005, Koo et al. [42] fabricated PtNPs without any additional capping agent (a material covered on the surface of NPs to avoid the aggregation) by exposing an atmospheric-pressure alternating-current (AC) $H_2$/He plasma on an aqueous solution of $H_2PtCl_6$ [Fig. 7(a)]. The transmission electron microscopy (TEM) image [Fig. 7(b)] indicates that the average size of PtNPs is about 2 nm and they demonstrate a nanocrystalline nature. The authors also found that the average size of PtNPs increased from 2 nm to 5 nm when the solution temperature changed from 10 to 40 $^oC$. In that paper, the authors concluded that the $PtCl_6^{2-}$ was reduced by atomic H in assistance of $H^+$, produced from the $H_2$/He plasma. This in part is correct, however, the $PtCl_6^{2-}$ should be reduced not only by atomic H, but also by the other capable reducing species mentioned in Section 2.1.1.

After the work of Koo et al. [42], noble metals of Au [43, 46, 67, 71, 73, 135-142], Ag [46, 69, 143-147], Pt [67], and Pd [148-153] NPs were intensively explored by using plasma over liquid configurations.

Sankaran's group [46, 69] obtained Au and Ag NPs from a DC microplasma-liquid system with an H-shape glass cell by reducing metal ions either originating from the solution or from anodic dissolution of Au (Ag) foil, fructose was present in solution to prevent agglomeration and precipitation of the particles, and they found that the NPs were formed just near the plasma cathode as shown in Fig. 8(a) in the anodic dissolution. They attributed the NPs formation to the plasma-provided-electron reduction of Au or Ag ions either from the anodic dissolution or the pristine metal salt solutions. Recently, a similar experiment has been performed by replacing the immersed metal anode with a plasma [154-156]. As shown in Fig. 8(b), AgNPs were formed just near the plasma cathode part from $AgNO_3$ solution, while there is no observable AgNPs in the plasma anode case although the plasma anode can provide reducing species as described above. The reduced Ag might be oxidized to $Ag_2O$ or AgO since the plasma also generates oxidizing species. Besides reaction $Ag^+ + e^- \rightarrow Ag$ ($E^o$=+0.80), reactions $2AgO + H_2O + 2e^- \rightarrow Ag_2O + 2OH^-$ ($E^o$=+0.20) and $AgCl + e^- \rightarrow Ag + Cl^-$ ($E^o$=+0.60) might also occur in the solutions since its negative sign of Gibbs free energy. It is interesting to note that AuNPs can be formed near both plasma anode and cathode when using an aqueous solution of $HAuCl_4$ in the same experiment [Fig. 8(c)]. The AuNPs is concentrated near the plasma anode while the AuNPs near the plasma cathode display a slight concentration gradient. Then a question arises: How do we understand the difference between the formation processes of AgNPs and AuNPs under the similar plasma conditions? We suggest that the nature of the metal ion in the solution might give some insights into this question. Ag ions in the solution take the form of $Ag^+$, while Au ions mainly take the form of $[AuCl_x(OH)_{4-x}]^-$ ($x \geq 2$ at low pH values, $x<2$ at high pH values) [91, 92]. Therefore, they will move in the solution with different directions under the same applied electric field. Consequently, AuNPs and AgNPs are formed near plasma anode and plasma cathode, respectively. Of course, there also exists a certain amount of $[AuCl_x(OH)_{4-x}]^-$ near the plasma cathode and $Ag^+$ near the plasma anode in spite of the electric field-driven movement of ions, while whether the reductions happen or not is determined by the electrochemical series of Au and Ag, the concentrations of Au and Ag ions, and the yields and types of species produced at plasma anode (mainly $Cl_2$ for $AgNO_3$) or cathode (mainly $H_2$ for $HAuCl_4$). From Table 1, $AuCl_4^-$ is easier to be reduced than $Ag^+$. In addition, the ion mobility for $[AuCl_x(OH)_{4-x}]^-$ is smaller than $Ag^+$, resulting in a slower concentration decrease of $[AuCl_x(OH)_{4-x}]^-$ at cathode than $Ag^+$ at anode by the applied electric field. Therefore, AuNPs were observed at both plasma electrodes, while AgNPs were found just at the plasma cathode. At present, there is lack of understanding in the influence of the metal ions' nature on the metal NPs synthesis process and quality of final products. Attention must be paid to this subject in the future work.

Recently, Chen et al. [71] synthesized AuNPs from an aqueous solution of $HAuCl_4$ with deoxyribonucleic acid (DNA) as capping agent by using a pulse plasma at a pressure of 20 KPa. The authors found that uniform-sized AuNPs can be formed rapidly when plasma was irradiated on the solution. DNA can acts as capping agent for the forming AuNPs by electrostatically conjugated on the surface of AuNPs. Two kinds of DNA were used based on their different adsorption energies on the Au surface, i.e., 30-mer DNA consisting of guanine or cytosine base (denoted as $dG_{30}$ or $dC_{30}$). The results implied that the size and assembly of AuNPs can be tuned by the DNA type and concentration as shown in Figs. 9(a) and (b), which subsequently modulates the spectroscopic properties of AuNPs, i.e., tailoring the LSPR of AuNPs. It is well known that there exist Coulomb repulsion and van der Waals attraction between the metal NPs in the solution [157]. The unbalance between these two forces can cause the NP growth by coalescence or NP self-assembly. The authors suggested that the DNA can tune the forces between the forming

AuNPs by conjugating on the surfaces of AuNPs. Depending on the concentration and type, the final size and morphology of AuNPs can be tuned as illustrated in Fig. 9(d).

Although the metal NPs synthesis from aqueous solutions by plasma is very convenient, the high vapor pressure of water makes it limited to use plasmas operated at pressures from several tens kPa to atmospheric pressure. Therefore, the plasma is limited to a small size which limits the yield of metal NPs. Thanks to ILs which have a very low vapor pressure, plasma-IL system can be operated at low pressure where traditional large size processing plasmas can be applied. Endres' group co-worked with Janek's group initially started the AgNPs synthesis by reducing $CF_3SO_3Ag$ dissolved in IL of 1-butyl-3-methylimidazolium trifluoromethylsulfonate ([BMIm]TfO) [44, 45]. Later, many works were followed to synthesize noble metal NPs from the plasma-IL system, such as the work from the groups of Endre's [58, 158], Hatakeyama's [43, 67, 148, 159-161], and Liu's [47, 139, 162]. On one hand, NPs synthesized from the plasma-IL system do not need any additional surfactant to stabilize, and the size can be easily tuned from the plasma parameters at low pressure. On the other hand, the conjugation of ILs on the surface of NPs is very difficult to be eliminated when one wants to change the surface function of NPs by exchanging the surface conjugated molecules. Most importantly, the solubility of many metal salts in the IL is lower than that in water, and to synthesize some metals, we need specific metal salt or specific IL to initiate the dissolution, and the derivatives produced from decomposition of IL by plasma might contaminate the formed NPs. The reducing species generated from plasma usually move more easily in water-based solutions than the IL-based ones. Thus, we should address the above critical issues before we can intensively exploit plasma-IL system for metal NPs synthesis.

Plasma in liquid is also used in synthesis of noble metal NPs. AuNPs have been synthesized from reducing $AuCl_4^-$ by plasma in aqueous solutions [48, 49, 163-166]. Without considering the other mechanisms, Hieda et al. [48], Sato et al. [49] and Bratescu et al. [164] just considered that H radicals produced by plasma acts as the reducing species. Interestingly, Sato et al. [49] investigated the dynamic of the AuNPs formation during the plasma ignition with time-resolved TEM. As illustrated in Fig. 10, the reduction of $AuCl_4^-$ by plasma forms a loose self-assembly (100 nm or more) of small Au nanoclusters (less than 1 nm) at the very first stage, perhaps due to the van der Waals attraction (not mentioned by the authors). In the second stage, with the discharge plasma time lapses, the pH value of solution decreases due to the major contribution of the $HNO_3$ formation from the interaction between plasma and $N_2$ as well as $O_2$ dissolved in the solution and the minor contribution of the water vapor decomposition ($2H_2O+e \rightarrow 2H_2+O_2+e$, $H_2+e \rightarrow 2H^++2e$) . At low pH values, the self-assembly of Au nanoclusters disassembles. Then the Au nanoclusters form spherical AuNPs (perhaps growth by coalescence) with size smaller than the previous assembly. The pH value decreases with increasing discharge time, and when the pH value is lower than a critical value, the surface Au atoms of formed spherical AuNPs is dissolved into the solution which has been confirmed by measuring the time evolution of $AuCl_4^-$ concentration, showing a valley-like shape. Finally, the size-decreased spherical AuNPs as well as anisotropic AuNPs are formed in the solution due to nonequivalent dissolution rate of different Au facet. In addition, based on the results of X-ray photoelectron spectroscopy (XPS), time-of-flight secondary ion mass spectrometry (ToF-SIMS), and ultraviolet-visible (UV-vis) absorption spectroscopy, Bratescu et al. [164] found that the surfaces of synthesized AuNPs are partially oxidized when synthesized in solution with pH value of 12, while the surfaces are surrounded with gold chloride compounds when synthesized in solution with pH value of 3.

Besides the usage of chemical processes for the NP synthesis in plasma-liquid system, using evaporation of metal wires by discharge plasmas in electrolyte liquid or water, AgNPs [122, 123] and AuNPs [123] were formed. It was found that the type of liquid medium or composition can influence the particles size and size distribution. Moreover, physical sputtering is also used in a plasma-IL system, Torimoto et al. [120, 121] first reported that AuNPs and Au/AgNPs can be formed by physical sputtering of a target Au or Au/Ag foil in a setup as shown in Fig. 5. After these inspiring works, the physical sputtering method flourished in the NPs synthesis [167-172]. The influences of sputtering conditions and the IL composition on the final products were also investigated. Hatakeyama et al. [173] found that the size of NPs is independent of the discharge time, pressure and current; and the electrodes' distance, while strongly related to the target and IL temperatures, and discharge voltage. Wender et al. [174] observed that the size and size distribution of NPs correlate with the IL structure, especially with the surface compositions of ILs, but independent of the surface tension or viscosity. This sputtering method is suitable for almost all metals and they can be used to synthesize metal NPs and alloys with tunable composition. But similar to the NPs synthesized by chemical processes in plasma-IL system, the resultant products might be contaminated with ILs, and this problem we must overcome in further works.

### 3.2. Magnetic Materials

Magnetic NPs, with both nanoscale size and magnetism, offer exciting opportunities for a wide range of applications, including data storage [175], magnetic fluids [176, 177], catalysis [178], magnetic resonance imaging [179, 180], and biomedicine [181, 182]. In last decades, various approaches have been developed to synthesize magnetic NPs, among them, the representatives are co-precipitation, thermal decomposition and/or reduction, micelle synthesis, hydrothermal synthesis, and laser pyrolysis techniques [183]. According to the low cost and the simplicity of experimental setup, the PLI method has recently demonstrated a promise to fabricate magnetic NPs with low cost and high speed.

NiNPs have been synthesized from discharge plasmas generated on a nickel cathode (immersed in $H_2SO_4$ [184], $K_2CO_3$ [123] and NaOH [125, 185] solutions), and the NiNPs were observed in the solution as well as on the surface of the Ni cathode. The scanning electron microscope (SEM) image of NiNPs in the solution synthesized by Toriyabe et al. [123] is shown in Fig. 11(a), indicating the NP size of hundreds nm with a relatively wide range distribution. Further investigation implied that the mean size of synthesized NiNPs decreased as the discharge voltage increased. Later, Akiyama et al. [185] tuned the size of NiNPs simply by the solution (NaOH) concentration since the applied discharge voltage strongly depends on the solution concentration at a constant power input. Moreover, if the cathode was covered by a quartz glass tube, the size distribution and the oxidation of synthesized NPs were much suppressed [Fig. 11(b)], and the authors attributed it to the partial discharge on the cathode caused by the cathode covering, which limits the overheating of the Ni cathode [125].

Using pulse discharge plasmas generated between tips of two corresponding Fe rods, micrometer $\gamma$-Fe synthesis has been reported in a liquid ammonia [186], and iron carbide NPs (orthorhombic $Fe_3C$ and monoclinic $\chi$-$Fe_{2.5}C$) encapsulated by multilayered graphite sheets were also formed in ethanol by adding Ar gas and an ultrasonic cavitation bubble into the plasma zone [187, 188]. In an experimental setup similar to Ref. [186], NPs of $Fe_3O_4$ [189] and onion-like carbon-encapsulated Co, Ni, and Fe (Co@C, Ni@C, and Fe@C) [190, 191] were fabricated from an aqueous solution of cationic surfactant (1-hexadecylpyridinium bromide monohydrate, CPyB) and ethanol, respectively. Indeed, this fabrication takes advantage of both anodic dissolution and cathode evaporation as mentioned in Section 2, and the yield must be enhanced in contrast to the process just using either anodic dissolution or cathode evaporation. The utilization of cationic surfactant and ethanol renders a surface modification, resulting in the tailoring of the size, stability and oxidation for NPs. Due to the coexistence of reducing and oxidizing species in the plasma-liquid interface, the resultants should be a mixture of pure metal and metal oxidation. The authors [189] also found that the $Fe_3O_4$ NPs (~19 nm) were a mixture of Fe, FeO and $Fe_3O_4$, while the purity of $Fe_3O_4$ increased with increasing CPyB concentration. A purity of 98 % for $Fe_3O_4$ NPs can be obtained in an aqueous solution of CPyB (0.84 g in 200 ml distilled water). TEM images of Fe@C and Ni@C synthesized from ethanol are shown in Figs. 11(c) and (d), respectively. Obviously, the synthesized NPs are coated with a few-layer of carbon which are formed from the carbon precursors produced by the plasma-induced ethanol decomposition. Interestingly, the results of X-ray diffraction (XRD) and energy-dispersive X-ray (EDX) spectroscopy for the NPs indicted that there was no observable oxidation for the core metal. The non-oxidization of NPs might be due to the presence of ethanol which can scavenge the plasma-induced strong oxidizing species of OH radicals and generate reducing species (H and $H_2$) as described in Eq. 6 [192, 193],

$$OH + C_2H_5OH \rightarrow H_2O + C_2H_4OH \quad (6.1)$$

$$e_{aq}^- + C_2H_5OH \rightarrow H + C_2H_5O^- \quad (6.2)$$

$$H + C_2H_5OH \rightarrow H_2 + C_2H_4OH \quad (6.3)$$

By encapsulating magnetic NPs in carbon, the NPs showed high coercivities: 296 Oe for Fe@C and 189 Oe for Ni@C, comparing to those of bulk Fe (90 Oe) and Ni (70 Oe) [191] (The unit of coercivity was mistaken as T in Ref. [191], we correct it as Oe in this review.). In addition, the results of human lung epithelial A549 cells exposure to Fe@C, Co@C and Ni@C NPs showed a low cytotoxicity of these NPs [190]. The high magnetic properties and low cytotoxicity renders these NPs possible manipulation in biomedicine application. Pure FeNPs (>100 nm) [194] or pure CoNPs (>90 nm) [195] were also reported to be generated in arc discharge over a liquid of ethylene glycol with two Fe or Co electrodes, where the ethylene glycol might play the same role as ethanol in refs. [190, 191]. Using molten salt LiCl–KCl–CsCl (over 300 °C) as the solution, Fe [196], Co [197], and Ni [198-200] NPs were synthesized from the dissolution of corresponding metal anodes under plasma exposure to the molten solution. However, the high processing temperature leads to broad size distribution. To control the size and size distribution, a disk anode was rotated, which can change the plasma exposure spot with rotation, decreasing the delivery quantity of reducing species onto the exposed spot [198]. As a result, smaller size and more uniform size distribution can be obtained in a higher speed rotation of anode.

Besides the consumption of the electrodes, the solution as a precursor is also used to synthesize magnetic NPs as in aforementioned case of the AuNP synthesis. Figure 11(e) shows rice-shaped NPs which were synthesized from aqueous solution of $FeCl_3 \cdot 6H_2O$ and polyethylene glycol with pulse plasma generated between two tungsten electrodes [201]. XRD pattern of these NPs indicated that the NPs consisted of ($\alpha$-$Fe_2O_3$) and iron oxide-hydroxide ($\beta$-FeO(OH). The formations of Co (0) [202] or Co (II) oxide [203] NPs were reported in experimental setups similar to Ref. [201] by only replacing the solution with aqueous solutions of $CoCl_2 \cdot 6H_2O$ and sodium dodecyl sulfate or $Co(CH_3COO)_2 \cdot 4H_2O$. While the conclusion for the formation of pure CoNPs needs to be rechecked since the reactivity of cobalt and the presence of strong oxidizing species.

$FeO_x$ NPs can also be rapidly synthesized by an atmospheric-pressure multi-microplasmas generated over an aqueous solution of $FeCl_2$ and dextrane and dimercaptosuccinic acid [204]. The plasma acted as a cathode, while a graphite rod immersed in the solution was used as anode. The as-synthesized $FeO_x$ NPs were purified by removing the excess of ions and ligand molecules with dialysis or size exclusion chromatography. Ultra-fine $FeO_x$ NPs (≤5 nm) with narrow size distribution were obtained after the purification process, as shown in Fig. 11(f). Rudimental magnetic resonance imaging assessments confirmed the "positive" contrast enhancement effect when the synthesized $FeO_x$ NPs were used as the contrast agent.

### 3.3. Cu

CuNPs with controllable shapes have shown significant catalytic [205], optical [102, 206], and conducting properties from which many fundamental and technological research might benefit. However, Cu is very reactive when its size decreases to nanometers, it is difficult to fabricate pure metallic CuNPs from solution-based process unless there exists very strong reducing agents and carefully selected protective capping material. In order to avoid possible oxidation from the ambient, strong reducing agent of hydrazine have been used in the synthesis of pure metallic CuNPs [207, 208].

As we know in Section 2, the PLIs can simultaneously generate strong reducing species as well as strong oxidizing species in aqueous solutions. One might consider that it is infeasible to synthesize pure metallic CuNPs from the PLIs. Actually, without or with unsuitable capping agents, many efforts on this subject produced partially oxidized CuNPs [58, 209-213]. However, the formation of CuNPs from the PLIs might be possible if we select appropriate capping agents and note the competition between the reducing and oxidizing species generated from the PLIs. The pursuit has already made progress by selecting suitable protective capping agents [214-216].

The CuNP synthesis have been performed from atmospheric-pressure AC plasma generated between two Cu rods (one over and the other in the aqueous solution) [214]. The authors found that the utilization of pure deionized water leads to a formation of spindle-like $Cu_2O/CuO$ nanostructures, while the solution of ascorbic acid or a mixture of ascorbic acid/cetyltrimethylammonium bromide (CTAB) results in the formation of metallic CuNPs, which was attributed to the protective effect of ascorbic acid on the nascent Cu seeds. Interestingly, as shown in Fig. 12(a), the surfactant CTAB showed a critical influence on the self-assembly of spherical CuNPs (with diameter of 700 nm-1 mm). In a reduced pressure (30 Torr), spherical CuNPs can also be formed from a pulse discharge plasma generated between two Cu rods immersed in pure ethylene glycol, while if deionized water was mixed with ethylene glycol, the products turned to needle-like CuO, polygon $Cu_2O$, and square $Cu_2O$ NPs depending on the water content.[215] The different morphologies were ascribed to the influence of different thermal conductivity of the used liquids on the formation process of these NPs.

Instead of consumption of the electrode material, the CuNPs formation from the PLIs by reduction of dissolved metal ions has also been reported [216]. An aqueous solution of ascorbic acid/gelatin/$CuCl_2$ was used as the Cu source, the NPs were formed from a pulse discharge plasma generated between two tungsten electrodes submersed in the solution. XRD results of the products indicated a decrease in oxidization with increasing discharge time. Based on a thermogravimetric analysis, the authors assumed that the gelatin played an important role in protecting CuNPs. As the discharge time increased, the gelatin fractions produced by plasma decomposition and the solution temperature were increased, while the pH value was decreased. Thus, the synthesized CuNPs were protected due to the ability of scavenging oxidizing species of the gelatin fractions [217] together with the relaxation of the random coil of gelatin [218] at low pH value and high temperature. Figure 12(b) gives the synthesized multi-shaped CuNPs. It is worth noting that the CuNPs are polydisperse rather than monodisperse colloids. Future work should pay attention to the formation of shape-controlled monodisperese CuNPs for advanced applications.

### 3.4. Metal Oxides

In contrast to the pure metallic NPs of reactive metals, the formation of corresponding metal oxides from the PLIs is relatively easy due to the existence of strong oxidizing species. Various kinds of metal oxide NPs fabricated from the PLIs have been reported [61, 219-234]. In the following, we will give two representative results for the PLIs induced metal oxides NPs.

An AC plasma was generated between a Cu filament and an aqueous solution of $NaNO_3$, and another Cu filament was immersed in the solution as the counter electrode [233]. Cu ions came from the anodic dissolution of the immersed Cu filament and the evaporation of the other Cu filament, and reducing species generated by the PLIs reduced these Cu ions to form NPs. Figure 13 gives the TEM images of products obtained from this AC plasma over liquid system. As shown in section 3.3, the oxidation state of final products depends on the available capping agents. Without any surfactant, the as-synthesized products were small CuO particles, and grew into rice-shaped CuO nanorods [Figs.13 (a) and (b)]. However, the small products grew into $Cu_2O$ or CuNPs when ascorbic acid or hydrazine hydrate was added into the $NaNO_3$ solution. One possible explaination is that the oxidizing species produced from the PLIs can be scavenged by the additives, and different ablities of scavengers result in different oxidation state of the product.

ZnONPs can also be formed from plasma in liquid [234]. A Zn wire (as cathode) and a mesh made from Pt wire (as anode) were submersed in an aqueous solution of $K_2CO_3$. When a certain voltage was applied, plasma was generated in the immersed part of Zn wire. A medium power input results in a formation of flower-like ZnONPs [Figs. 14(a) and (b)], while high power input leads to a formation of aggregated ZnONPs [Fig. 14(c)]. Figure 14(d) presents the mechanism of the synthesis. The Zn wire is oxidized to ZnO on its surface after plasma generation. At low power input, the surface is under the melting point and some $Zn(OH)_4^{2-}$ can be formed around the Zn wire by $ZnO(s) + H_2O + 2OH^- \rightarrow Zn(OH)_4^{2-}$, and then the $Zn(OH)_4^{2-}$ transfers to the low temperature zone and decomposes into ZnO by $Zn(OH)_4^{2-} \rightarrow ZnO(s) + H_2O + 2OH^-$. Theses ZnO grow preferentially along the [0 0 0 1] direction [235, 236] to form flower-like ZnONP. However, at high power input (>300 W/cm$^{-2}$, data taken from Ref. [234]), the temperature of Zn wire reaches its melting point. Rapid evaporation supplies Zn and ZnO, resulting in the aggregation of ZnO particles, as shown in Fig. 14(c).

### 3.5. Alloy

It is well known that NMs usually lack of stability due to their large volume fraction of grain boundaries. In addition to the surface functions by various ligands, alloying of NMs is one alternative strategy for stabilizing NMs [237, 238]. The metal can come either from the electrode materials or from the working solutions. This way, the PLIs can also be used to fabricate alloying NMs.

A displacement reaction has been used to synthesize FePt [196, 239]. FeNPs was first formed from an Ar plasma-induced cathodic discharge generated over a molten $FeCl_2$/LiCl-KCl-CsCl (573-773 K), and then $PtCl_2$ was added into the FeNPs dissolved molten solution, continuous plasma treatment leaded to the displacement reaction of $2Fe+Pt^{2+}\rightarrow FePt+Fe^{2+}$, then a process of later solid-phase interdiffusion resulted in the formation of FePt (~100 nm). In fact, the XRD data indicated that the first-step synthesized FeNPs were actually dominated by $Fe_2O_3$ phase and the final products were a mixture of Fe/Pt/FePt/ $Fe_2O_3$. The authors suggested that the oxidation of the FeNPs confirmed by XRD data was probably by the water washing process during the extracting period. Presumably pure $Fe_xPt_y$ NPs can be obtained by optimizing the ratio of $FeCl_2$ to $PtCl_2$ and discharge time. The synthesis of CoPt [197, 198], $CoPt_3$ [197], and SmCo [240] NPs were also reported by a similar method.

By generating a DC plasma between an $Fe_{50}Pt_{50}$ plate anode and an Fe rod cathode (both immersed in ethanol), FePt alloy NP with a broad size distributions from several to several hundred nm were synthesized [241]. An ultrasonic system was used to produce cavitation bubbles between the electrodes for facilitating the discharge. Therefore, the resultant products were formed by plasma discharge in liquid ethanol with an ultrasonic cavitation field. Based on the XRD and TEM analysis, the authors concluded that the products were carbon encapsulated and most of the cores had average composition of paramagnetic Pt($Fe_{38}Pt_{62}$). Figure 15 presents the hysteresis loops of the (A) as-synthesized and (B) 923 K annealed carbon encapsulated Fe-Pt alloy NPs. For the as-synthesized product, the coercivity is 6.4 $kAm^{-1}$ and the magnetization saturation is 16.7 $Am^2kg^{-1}$ at room temperature. While the coercivity and the magnetization saturation change to 73.6 $kAm^{-1}$ and 13.5 $Am^2kg^{-1}$, respectively, after 923 K annealing. The small decrease of magnetization was accounted for possible redistribution of chemical elements (Fe-Pt) during the annealing (toward Pt-rich composition), and the coercivity change was related to the conversion of the disordered fcc structure to the ordered fct structure confirmed by XRD analysis.

The bimetallic Ag/Pt NPs were also reported to be formed by performing unipolar pulse plasma between a Ag cathode and a Pt anode in an aqueous solution of sodium dodecylsulfonate (SDS, 10 mM) and NaCl (0.1 M) [242]. Reducing species produced by the plasma reduce the Ag and Pt ions coming from electrodes, resulting in a formation of Ag-Pt NPs in the solution. Figure 16 (a) shows a TEM image of the Ag-Pt NPs fabricated by 30-s plasma treatment, and the NPs have a narrow size distribution with an average size of 5 nm. High-resolution (HR) TEM image and selected area electron diffraction (SAED) patterns of the synthesized Ag-Pt are presented in Figs. 16(c) and (d). These HRTEM and SAED results indicate that the bimetallic NPs were in the form of an immiscible nanocomposite with a distinct boundary between phases of Ag and Pt. Thus, the synthesized NPs were Ag/Pt bimetallic NPs rather than a usual alloy. Similar to this work, wurtzite-type ZnMgS synthesis was reported from a pulse plasma generated between two ZnMg rod electrodes inside liquid sulfur [243].

In addition to using the chemical processes in the PLIs, the physical sputtering process can also be used to synthesize alloy NPs, such as AuAg [121, 244, 245] and AuPd [246] by using the experimental setup as shown in Fig. 5. The alloy can be formed by sputtering a multi-metal-composed target with Ar plasma at low pressure, the sputtered metal clusters are collected by the IL just below the target. The IL provides a nucleation and growth field for the clusters and also acts as a stabilizing agent for the formed NPs. Because the target can be changed simply, it is potential to extend this method to the synthesis of various metal alloys.

### 3.6. Semiconductors

Since semiconductor quantum dots (QDs) were discovered by Ekimov et al. [247] in the early 1980s, they have been attracting much attention according to their size and shape dependent electronic and optical characteristics. These unique properties render QDs versatile applications such as in transistors, solar cells, LEDs, diode lasers, medical imaging, and quantum computing [248-255]. Stimulated by these promising applications, many approaches have been developed to synthesize the semiconductor QDs [256, 257].

In principle, it is also possible to form the semiconductor QDs by the PLIs. However, there are just a few reports on the semiconductor QD synthesis by the PLIs. The reason is partially due to the fact that it is difficult to find suitable liquid precursors containing semiconductor elements, for example, many Si-containing liquids are instable and sometimes toxic. Therefore, the suitable precursors are limited.

Silicon electrodeposition on a substrate has been reported by using Si-containing organic solutions [258, 259], ILs [260, 261], and molten salts [262-264]. Thus, semiconductor QDs can also be formed if the substrate in the electrodepositon system were replaced by a gaseous plasma. In fact, an effort has been made to synthesize Si and Ge [58, 265, 266] QDs in a low-pressure plasma-IL system, and TEM images of the synthesized Si and GeQDs from three types of ILs are presented in Fig. 17. The results indicated that the size for Si and Ge QDs are similar when using 1-ethyl-3-methylimidazolium bis(trifluoromethylsulfonyl)imide ([EMIM]$Tf_2N$), while the size of GeQDs is dramatically decreased by using 1-Ethyl-3-methylimidazolium tris(pentafluoroethyl)trifluorophosphate ([EMIM]FAP). The authors did not explain this phenomenon. Probably, the anions of $Tf_2N$ and FAP can serve as capping ligands[267] for the synthesized QDs, and this assumption can be tested by an EDX mapping of the QDs.

Strangely, there is no report on the SiQDs synthesis by the PLI-induced reduction of Si-containing aqueous solution, although there exists strong reducing species with $E^o$ as high as -2.87 V (Table. 1) in the PLIs. Because the strong reducing species, reactions such as $SiF_6^{2-} + 4e^- \rightarrow Si + 6F^-$ ($E^o$ =-1.37 V [268]) and $SiO_2 + 4H^+ + 4e^- \rightarrow Si + 2H_2O$ ($E^o$ =-0.99 V [90]) are possible in the PLIs, and SiQDs might be formed from these reactions. In addition, the IL is possible to be used as a flexible substrate for synthesizing SiQDs in low-pressure plasma, since the Si sputtering [269, 270], and the decomposition of Si-containing gas or liquid precursors [271-276] are feasible.

### 3.7. Novel Carbon-Related Materials

In the early 1990s, a method based on arc discharge induced carbon evaporation was developed for producing $C_{60}$ [277] in He and multi-walled carbon nanotubes (MWCNT) [278] in Ar. This method was later modified to produce a variety of polycyclic aromatic compounds [51, 279], and $C_{60}$ [51] by generating arc discharge plasma between two graphite [279] or pyrographite [51] electrodes inside liquid toluene. It was believed that the precursors used to form the products originated from the graphite electrode erosion for a low DC discharge voltage (24-28 V) [279], while for a high AC discharge voltage (10-20 kV), the presursors seemed likely from both the decompostion of toluene and the erosion of the electrode [51].

Stimulated by the unique poperties and a great number of potential applications of novel carbon-related NMs, the simple method of discharge plasma in liquid was used in many experiments in last decades, by just modifying some part of the prototype, successes have been made in producing MWCNTs [280-283], carbon onions [52, 284-286], carbon nano horns [287, 288], carbon NPs [117, 118, 289-295], and graphene nanosheets [117, 296-299]. One example of these fabrications is presented in Fig. 18. Two graphite electrodes were submerged in deionized water, and a discharge voltage of 16-17 V (30 A) was applied to these electrodes. Discharge plasma was generated in water by decreasing the eletrodes' gap to certain distance as shown in Fig. 18(a). Figures 18(c) and (d) present the low- and high-resolution TEM images for the products floating on the water surface. Obviously, the products consists of spherical carbon onions as well as polyhedral, nested onion-like particles. Since the water was used as the discharge medium, the products must be formed from the precusors produced by the interactions between plasma and electrodes. The synthesized novel cabon onions might be beneficial for potential appplictions such as solid-state lubrication [52] and supercapacitors [53].

More recently, nanodiamonds have been obtained from an atmospheric-pressure microplasma operated in gaseous phase of $Ar/H_2$/ethanol [300]. This result suggests that it is possible to fabricate nanodiamonds from microplasma generated over or inside liquid ethanol, while there is no relevant report so far. Therefore, we should pay attention to the microplasma-liquid ethanol system, perhaps nanodiamonds can also be found in it when optimizing the plasma parameters.

## 4. Perspective on the Future Research

The prospect of the NM synthesis by the PLIs is promising and exciting as they offer a rapid and easily controllable synthesis for a variety of NMs. However, there are several issues that must be addressed in the future work. They include: (1) size- and shape-controlled synthesis, (2) oxidization prevention for reactive metals, and (3) removal of impurity materials and large-scale production.

### 4.1. Size- and Shape-Controlled Synthesis

The physical and chemical properties of NMs are closely dependent on their sizes and shapes, and therefore size- and shape-controlled synthesis of NMs is great of importance [39]. Reports on the size-controlled synthesis of NMs from the PLIs have been presented [49, 72, 143, 164, 185, 301-304], while the shape-controlled synthesis is scarce, and even the few reported shape-controlled syntheses lead to a mixture of NMs with several shapes [49, 136]. To advance this research field, the size- and shape-controlled synthesis of NMs from the PLIs is necessary. We can address this subject from two directions: the liquid and the plasma.

Solution-based NM syntheses have left us ample strategies for size- and shape-controlled synthesis. The formation of NPs in solution is illustrated in Fig. 19 [305] based on the theory of V. K. LaMer and R. H. Dinegar [306]. The solute is the feedstock for NPs, say, $Au_n$ ($n \geq 1$) formed by reducing an aqueous solution of $HAuCl_4$ with a reductant. The solute concentration builds up as the reduction reaction proceeds. When the solute concentration reaches a critical value, the system becomes heterogeneous by a process of nucleation. The rate of nucleation depends on the solute concentration at which the system starts to nucleate. The number of formed nuclei increases with time and consequently resulting in rapid decrease of solute concentration. When the solute concentration is below the critical concentration, nucleation stops and the NP growth starts. If there is no surfactant, the formed NPs will finally aggregate to reduce the free energy of the system. This mechanism provides us a route to obtain size-controlled monodisperse NPs. If the amount of precursors is constant, the final size of formed NPs is related to the nuclei number, which can be changed by tuning the nucleation time from the reducing speed and the reaction temperature.

It is well known that the rate of chemical reaction is related to the temperature, pH value etc. Thus, the rate of nucleation might be tuned by the reaction environment such as reactants ratio, temperature and pH value. In fact, temperature effect has been shown in a $H_2$/He plasma- aqueous $H_2PtCl_6$ solution [42] and by sputtering gold electrode with discharge plasma in liquid nitrogen and water (0 °C) [302, 307]. The former case showed that the sizes of synthesized PtNPs were 2 and 5 nm at temperature of 10 and 40 °C, respectively. Similarly, the sizes of formed AuNPs in the latter case were ~1.5 nm in liquid nitrogen and ~ 3.5 nm in water, and these sizes are obviously smaller than those in the similar works performed at room temperature (sizes usually >10 nm). In addition, the size and assembly of NPs synthesized from the PLIs can also be tuned by the selection of surfactant type and concentration [71].

Seed-mediated [41, 308], plasmon-mediated [309], and template-based [310-312] growths of anisotropic NMs are popular techniques in solution-based processes. Electrochemical method also shows the possibility for synthesizing gold nanorods [102, 313, 314]. It is found that surfactants play an important role on adjusting the growth rate of different facets for the NMs, for instance, cetyltrimethylammonium bromide (CTAB) usually exists in the gold nanorod synthesis [315]. Thereby, we can lend these precious strategies to synthesize anisotropic NM from the PLIs, since the PLI method is just a modified solution-based system by replacing the reducing agents with the PLI-induced species. Moreover, the formation of gold nanorods from the PLIs [160] by using CNTs as templates proves the feasibility of these concepts.

Different from solution-based synthesis, the plasma-induced reducing species are usually not one type, but several more as discussed in Section 2. The difference of reducing ability for the species in the PLIs renders us a way to tune the nucleation and growth of NMs by varying the plasma parameters. For example, in a plasma over $HAuCl_4$ dissolved IL system, when the negative voltage was applied to the IL, small-sized AuNPs were obtained, while large-sized AuNPs were formed as negative voltage was applied to the counter electrode [159]. It was believed that the yields of reducing species induced by the PLIs were changed by changing the voltage polarity. The discovery of the short- and long-lived reducing species in a plasma-aqueous solution system [71] induces us to optimize the discharge time, by which the rates of nucleation and growth as well as the after plasma ageing growth might be controlled. Our own work also indicated that shape of the synthesized AuNPs can be tuned by discharge time in a plasma over an aqueous solution of $HAuCl_4$/DNA (not published, 30-mer DNA consisting of guanine base). Apart from the discharge time and voltage, other parameters such as gas atmosphere and pressure can also be used to adjust the yields of reducing species. For example, if we use $N_2/H_2$ or ammonia as the working gas, ample atomic hydrogen and certain hydrazine ($N_2H_4$) [97] will be produced by the plasma, the components of reducing species will be different from an Ar plasma-liquid system. As a result, the nucleation and growth of NPs in the solution will be changed by changing the gas atmosphere. At present, there is still a large demand for plasma-parameter-induced size- and shape-controlled synthesis of NMs, which will become of great interest in the future.

### 4.2. Oxidization Prevention for Reactive Metals

As described in Section 2, the interactions between plasma and an aqueous solution can produce reducing species, and in the same time oxidizing species. Therefore, in order to synthesize NPs of pure reactive metal such as Fe, Cu and Zn, oxidization in the process should be prevented in some methods. The surface coating by inertial material is one simple means, by which the coated NPs become oxidants resistant, stable and safe if used in vivo. The coating can be carried out either by using carbon-rich liquid or carbon electrode in the PLIs [190, 191]. It is also possible to protect the NPs by eliminating the oxidizing species in the PLIs with an oxidizing species scavenger either produced by the PLIs themselves or from an outside additive. Moreover, the stability of NPs can be increased by alloying them with other metals [237, 238, 316-318], and at the same time, specific properties such as strong ability in catalysis are followed. We should point out that it is possible to synthesize NPs of pure reactive metals if the working gas for discharge plasma is oxygen free, and the liquid is oxygen free or a molten metal salt, but the synthesized NPs are still very susceptible to oxidation from oxygen in the surrounding when they are transferred.

### 4.3. Removal of Impurities and Large-Scale Production

Impurities can enter the final products during the NM synthesis from the PLIs in several ways. If NMs are formed from the metal ions in solution, the undesirable impurities may come from the electrode by chemical reaction, physical sputtering or evaporation. The parameters for controlling these effects usually are specific for individually experimental setup, and thereby, requires extensive study to find optimal conditions to prevent or decrease the impurities based on the physics and chemistry underlying the plasma-liquid system.

It is suggested that preliminary attempt should be put on the electrode design. Coating the electrode with an inertial material such as carbon or quartz might be useful to prevent some liquid-based chemical reactions at the electrode, while this method must use pulse, radio frequency power or DC source with high voltage. On the other hand, the impurities from the electrodes can be excluded if an electrodeless discharge plasma (such as inductively-coupled plasma) is used.

The formation of NMs from the PLIs containing surfactants is inevitably accompanied by a decomposition of surfactants. The resulting fragments of surfactants as impurity might have positive or negative effect on the NM formation, which should be carefully studied.

In addition, plasmas used in the PLIs are usually operated at atmospheric pressure or several tens of kPa, the sizes of plasmas are limited to several mm at least in one dimension due to Paschen's law. For large-scale production of NMs from the PLIs, the plasma area should be increased. The recently developed plasma arrays [22, 319-324] might be a good choice since the plasma area is much increased by integrating many microplasmas into one array. We can also use other large-area atmospheric-pressure plasmas [325-333] which are based on different designs. Although the plasma area can be relatively large in a plasma-IL system at low pressure, the dissolution rates for most metal salts are smaller in IL than those in aqueous solutions, which limits the application of the plasma-IL system in the large-scale production of NMs.

## 5. Conclusions and Outlook

In this review, we have summarized the recent advances and present conditions of the NM synthesis from the PLIs. A theoretical analysis for the NM synthesis process is presented by analyzing the experimental data. Besides the theoretical analysis, the practical applications in several NM syntheses of the PLIs are also presented.

A discharge plasma contacting with a liquid provides a plasma-liquid interface where many physical and chemical processes take place. These processes includes decomposition, sputtering, and evaporation of liquid and the electrode material, producing complicated reactants including reducing as well as oxidizing species, and NMs can be formed by reactions between the reactants and the precursors originating from either the liquid or the electrodes. In particular, reducing and oxidizing species produced by the plasma-aqueous solution interactions are summarized: there exist long- and short-lived reducing species and their reducing abilities are roughly in scale with their standard reduction potentials. These findings indicate that it is possible to control the NM nucleation and growth by tuning the yields and ratios of yields for different reducing species. Besides the utilization of chemical processes, physical processes, such as the electrodes' sputtering and evaporation, are also presented for the PLIs. Moreover, a wealth of data is given for the PLI applications on the NM syntheses ranging from noble metals to graphene nanosheets. The solid experimental data and theoretical analysis suggest that the PLI method is a fascinating alternative for synthesizing various NMs.

To improve the application of NMs synthesized from the PLIs, the physical and chemical properties of the NMs should be tuned following different requirements from various applying environments. This target can be achieved by tuning the size and shape of NMs. Because the PLIs provide a larger parameter space compared with the solution-based synthesis. That is, the solution and the plasma, from which we can tailor the size and shape of NMs by tuning the nucleation and growth processes. Tailoring of plasma parameters allows us to control the yields of different reducing species, and then controlling the final size and shape of NMs.

Although many achievements on the PLI-induced synthesis of NMs have been achieved to date, the spatial and temporal evolutions of detailed processes taking placing in the solution, the plasma-liquid interface as well as the bulk plasma are still not well understood. Therefore, in-situ probing is needed to disclose the underlying secrets of the PLI-induced synthesis. After knowing the detailed processes, we might elaborately tune the specific reactions in the PLIs by changing the corresponding parameters, and finally obtain NMs with desirable size and shape.

Furthermore, due to the past work mainly focused on the synthesis of NMs from the PLIs, more detailed studies need to be implemented into the monodisperse synthesis and the property diagnostics for advanced applications of synthesized NMs in catalysis, solar cell, biomedicine etc.

ACKNOWLEDGMENTS

This work was partially supported by National Natural Science Foundation of China (Grant Nos.: 11405144, 11304132, 61376068, and 21322609), the Science Foundation Research Funds Provided to New Recruitments of China University of Petroleum, Beijing (QZDX-2014-01), and Thousand Talents Program. Q.C is greatly indebted to Dr. Marc-André Fortin of Université Laval and Dr. Naoki Shirai of Tokyo Metropolitan University for their kind permissions to reprint their TEM images.

**Table 1.** Standard reduction potentials ($E^o$) vs SHE for partially plasma-liquid interaction induced reactive species and some metal ions.

| Reactive species | Reactions | $E^o$ (V) | Ref. |
|---|---|---|---|
| Free electron | | N/A | |
| Secondary electron | | N/A | |
| Hydrated electron | $H_2O + e^- \rightarrow e_{aq}^-$ | -2.87 | [88] |
| H | $H^+ + e^- \rightarrow H$ | -2.30 | [88] |
| $H^-$ | $H_2 + 2e^- \rightarrow 2H^-$ | -2.40 | [334] |
| Hydrazine | $N_2 + 4H^+ + 4e^- \rightarrow N_2H_4$ | -1.21 | [335] |
| $H_2$ | $2H^+ + 2e^- \rightarrow H_2$ | 0.00 | [90] |
| $H_2O_2$ | $O_2 + 2H^+ + 2e^- \rightarrow H_2O_2$<br>$H_2O_2 + H^+ + e^- \rightarrow OH + H_2O$<br>$H_2O_2 + 2H^+ + 2e^- \rightarrow 2H_2O$<br>$H_2O_2 + 2OH^- \rightarrow O_2 + H_2O + 2e^-$ | +0.70<br>+0.96<br>+1.76<br>+0.15 | [334]<br>[334]<br>[90]<br>[77] |
| $O_2$ | $O_2 + 4H^+ + 4e^- \rightarrow 2H_2O$ | +1.23 | [77] |
| $O_3$ | $O_3 + 2H^+ + 2e^- \rightarrow O_2 + H_2O$ | +2.08 | [336] |
| O | $O + 2H^+ + 2e^- \rightarrow 2H_2O$ | +2.43 | [337] |
| OH | $OH + H^+ + e^- \rightarrow H_2O$ | +2.85 | [338] |
| $Au^{3+}$ | $Au^{3+} + 3e^- \rightarrow Au$ | +1.50 | [334] |
| $AuCl_4^-$ | $AuCl_4^- + 3e^- \rightarrow Au + 4Cl^-$ | +1.00 | [339] |
| $Ag^+$<br>$Ag^{2+}$<br>AgCl<br>AgO | $Ag^+ + e^- \rightarrow Ag$<br>$Ag^{2+} + e^- \rightarrow Ag^+$<br>$AgCl + e^- \rightarrow Ag + Cl^-$<br>$2AgO + H_2O + 2e^- \rightarrow Ag_2O + 2OH^-$ | +0.80<br>+2.00<br>+0.22<br>+0.60 | [77]<br>[340]<br>[77]<br>[340] |
| $Cu^{2+}$ | $Cu^{2+} + 2e^- \rightarrow Cu$ | +0.40 | [90] |
| $Fe^{2+}$ | $Fe^{2+} + 2e^- \rightarrow Fe$ | -0.44 | [90] |
| $Mg^{2+}$ | $Mg^{2+} + 2e^- \rightarrow Mg$ | -2.36 | [90] |
| $Cs^+$ | $Cs^+ + e^- \rightarrow Cs$ | -3.03 | [90] |

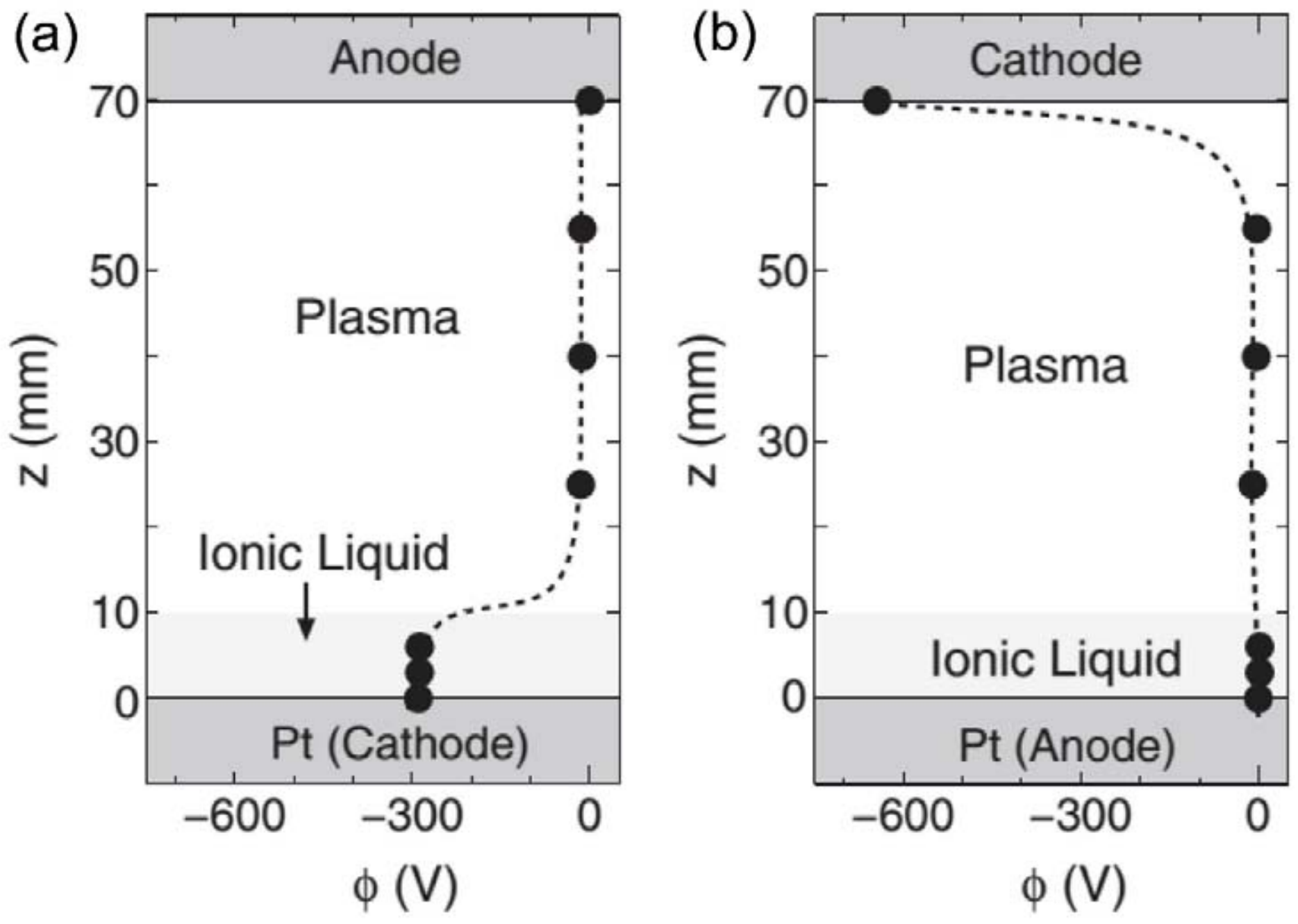

**Figure 1.** Plasma potential distributions along the axis of the electrodes for liquid as (a) cathode, and (b) anode. Reprinted with permission from Ref. [67]. Copyright 2009 WILEY-VCH Verlag GmbH & Co. KGaA, Weinheim.

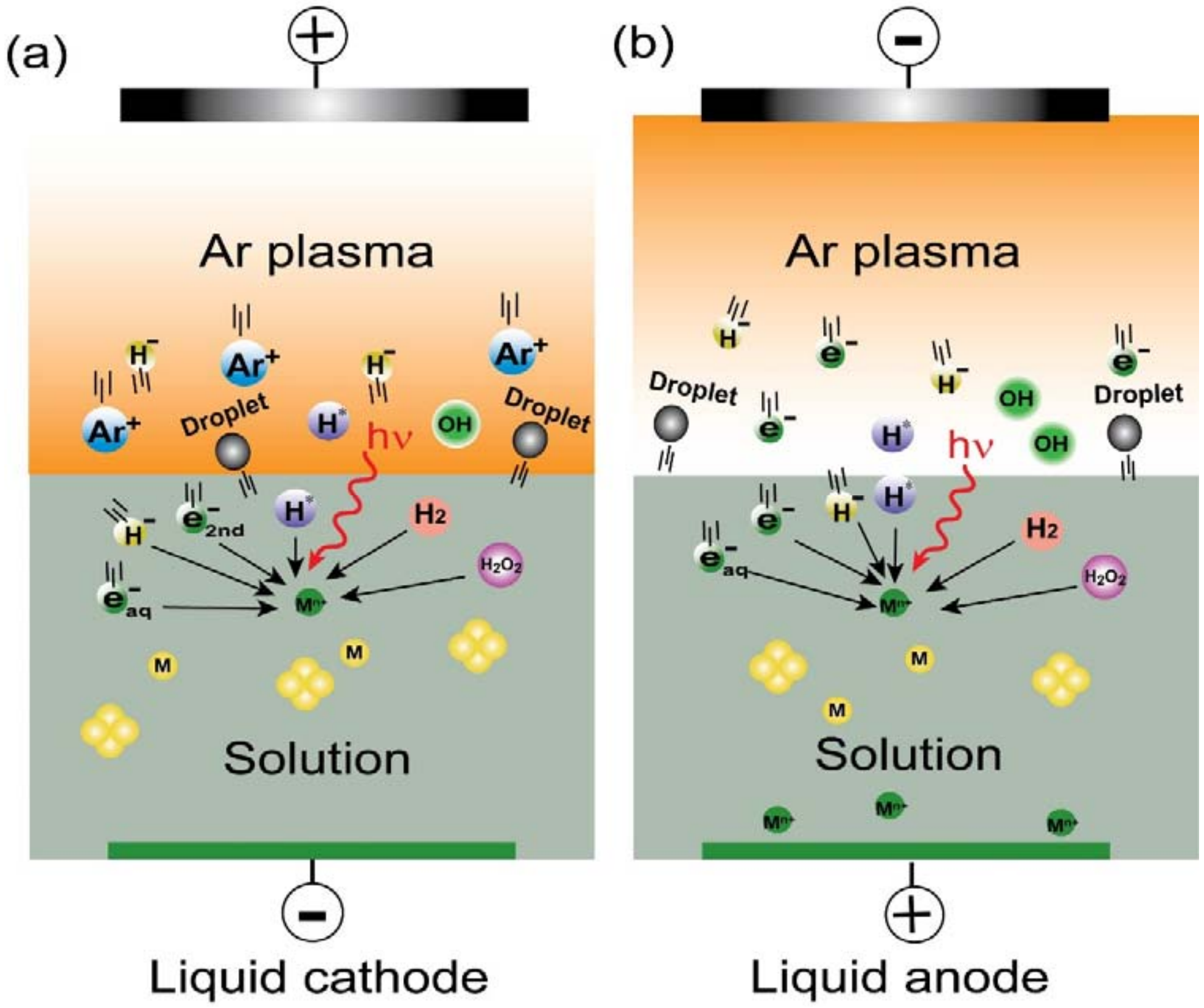

**Figure 2.** Physical and chemical processes in NM synthesis from plasma-liquid interactions. Liquid acts as (a) cathode, and (b) anode. Modified with permission from Ref. [55]. Copyright 2012 The Japan Society of Applied Physics.

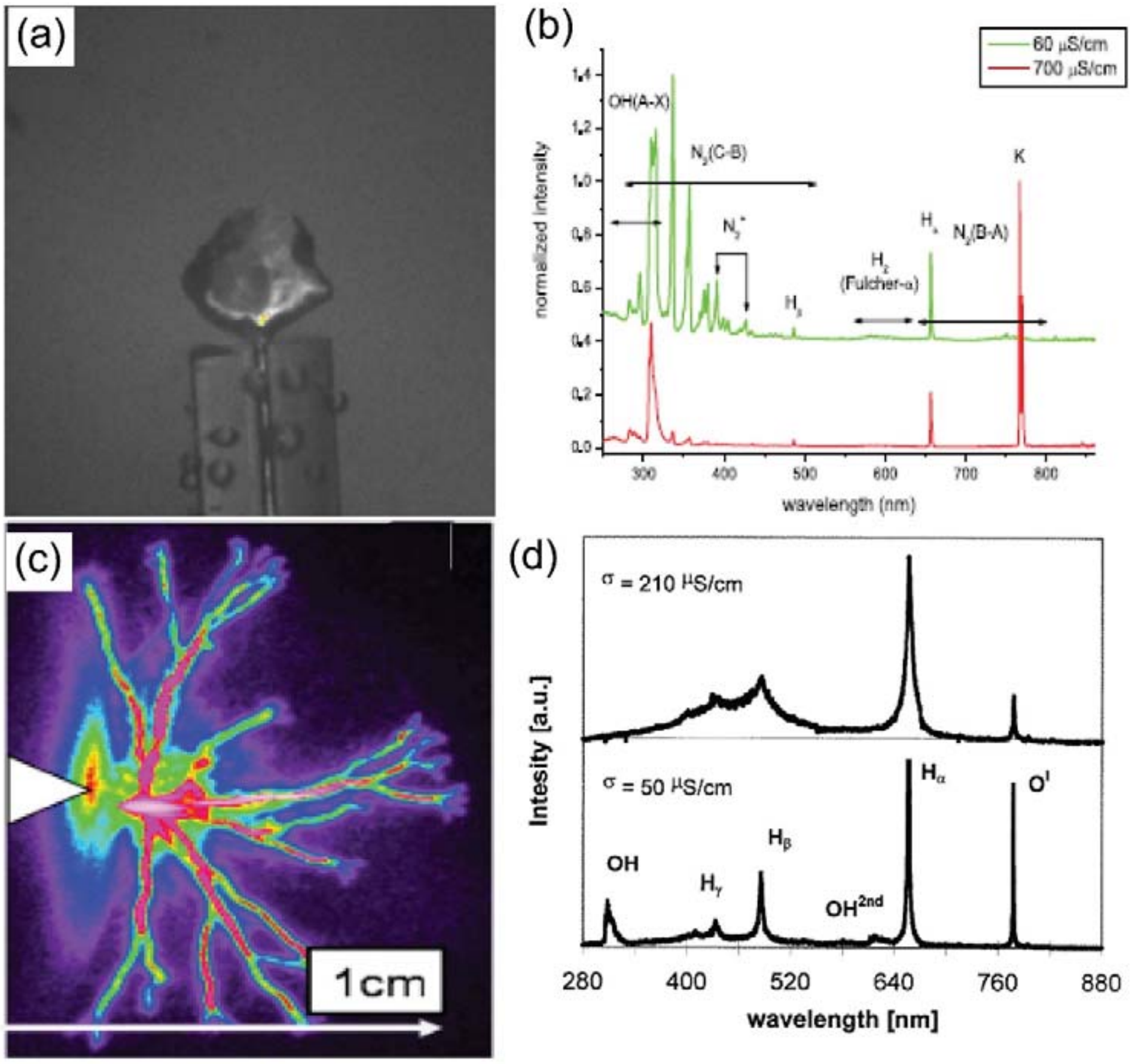

**Figure 3.** Photos and optical emission spectra for plasma in water bubble (a) and (b), and for streamer plasma in water (c) and (d). (a) and (b) are taken from Refs. [112] (c) and (d) are taken from Ref. [341] and Ref. [342], respectively. Reprinted with permissions from Refs. [112, 341, 342] Copyrights 2009 [112], 2011 [341], and 1999 [342] IOP Publishing.

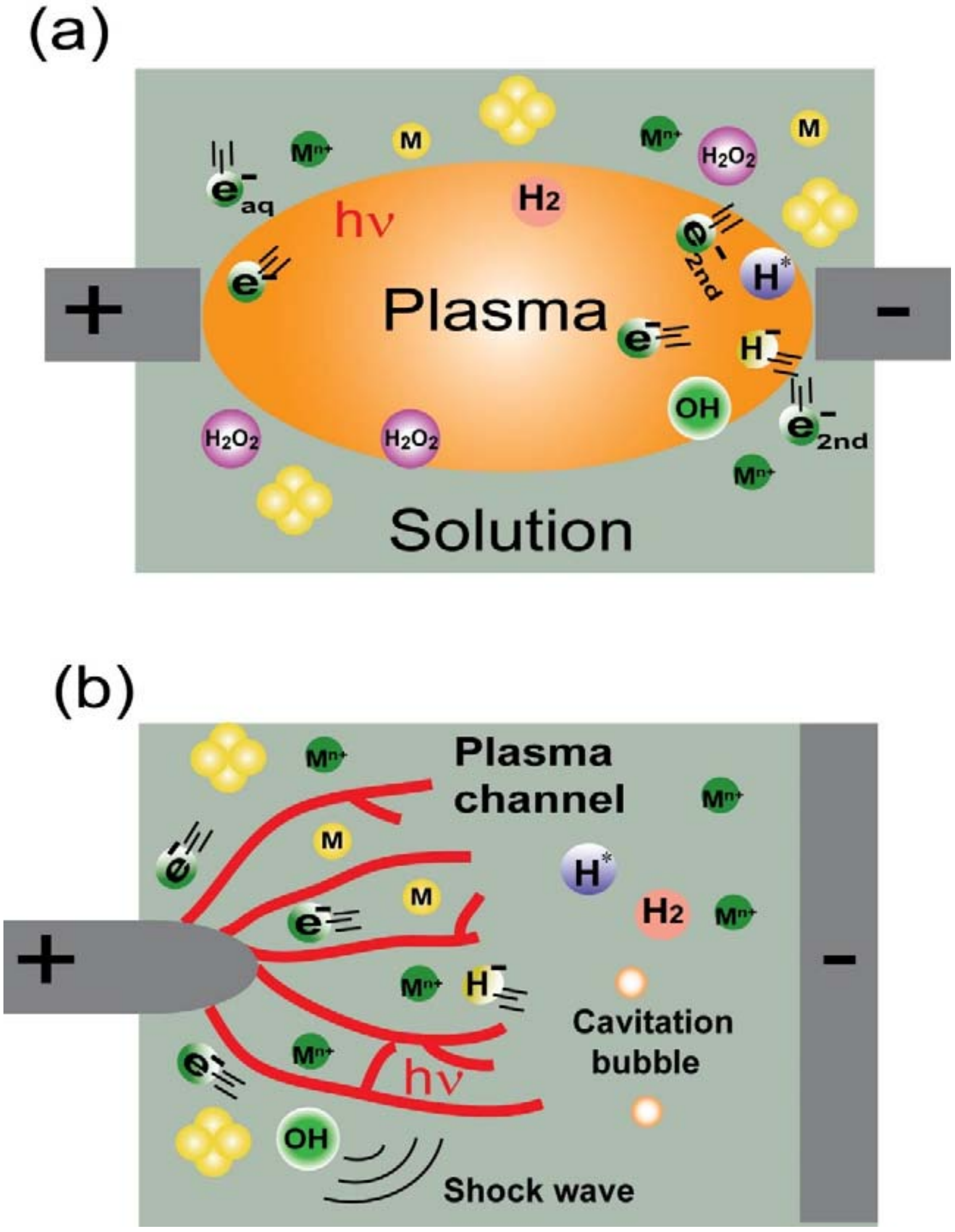

**Figure 4.** Physical and chemical processes in the NM synthesis from (a) plasma in bubble, and (b) streamer plasma in liquid.

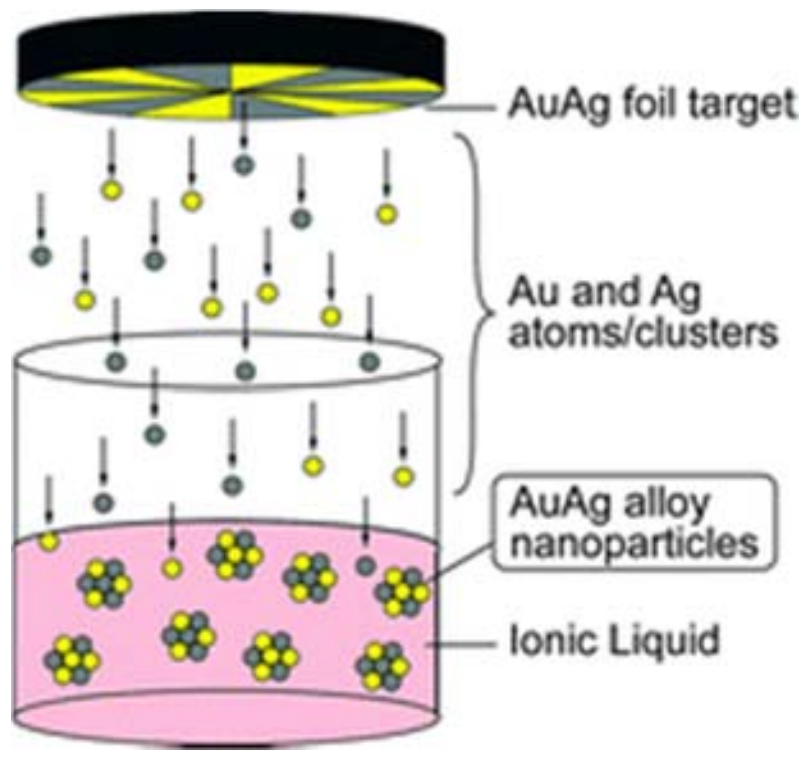

**Figure 5.** AuAg nanoparticles formation by sputtering AuAg in a plasma ionic liquid system. Reprinted with permissions from Ref. [121]. Copyright 2008 RSC publishing.

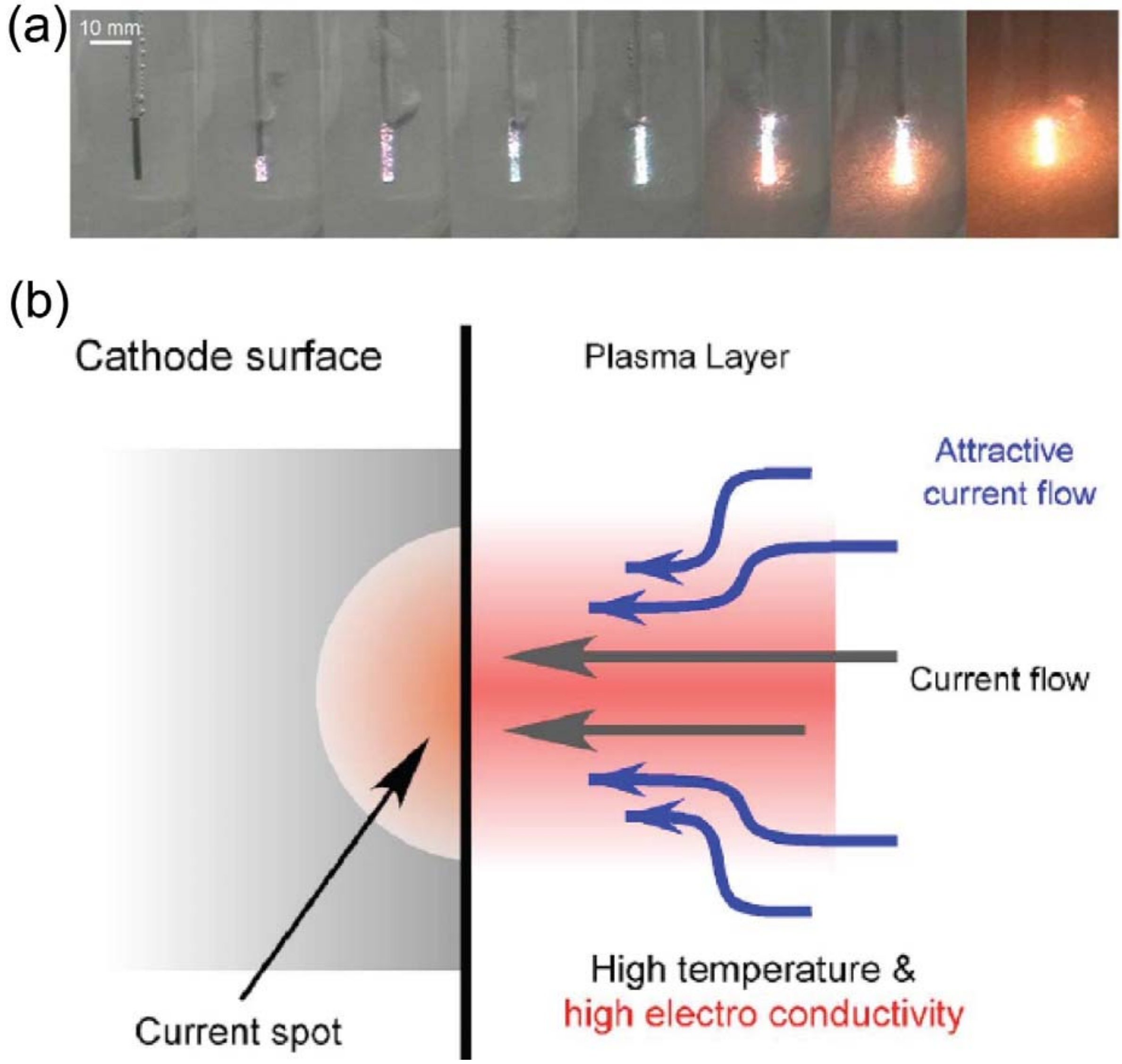

**Figure 6.** Metal nanoparticles formation by evaporation of electrode material. (a) Photos of plasma evolution in electrolyte solution under certain voltage, and (b) mechanism of the evaporation for metal electrode. Modified with permissions from Ref. [123]. Copyright 2007 American Institute of Physics.

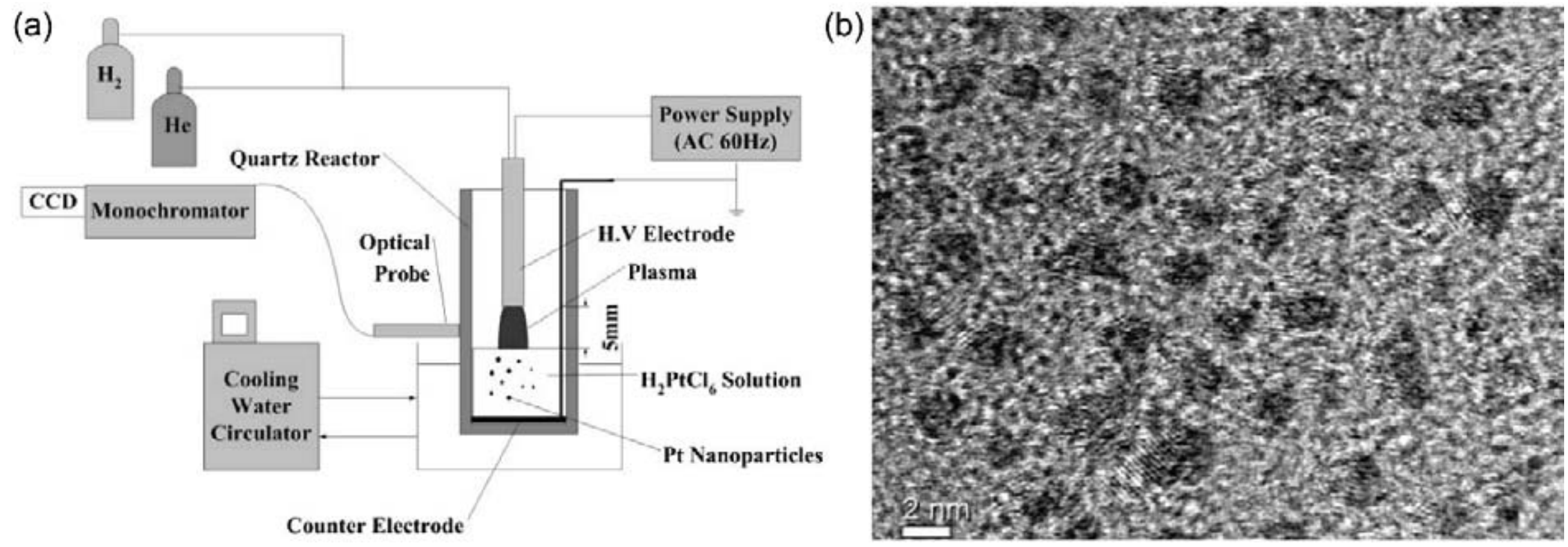

**Figure 7.** (a) Experimental setup for plasma-liquid synthesis of Pt nanoparticles, and (b) a representative TEM image of the synthesized Pt nanoparticles. Modified with permissions from Ref. [42]. Copyright 2005 The Royal Society of Chemistry.

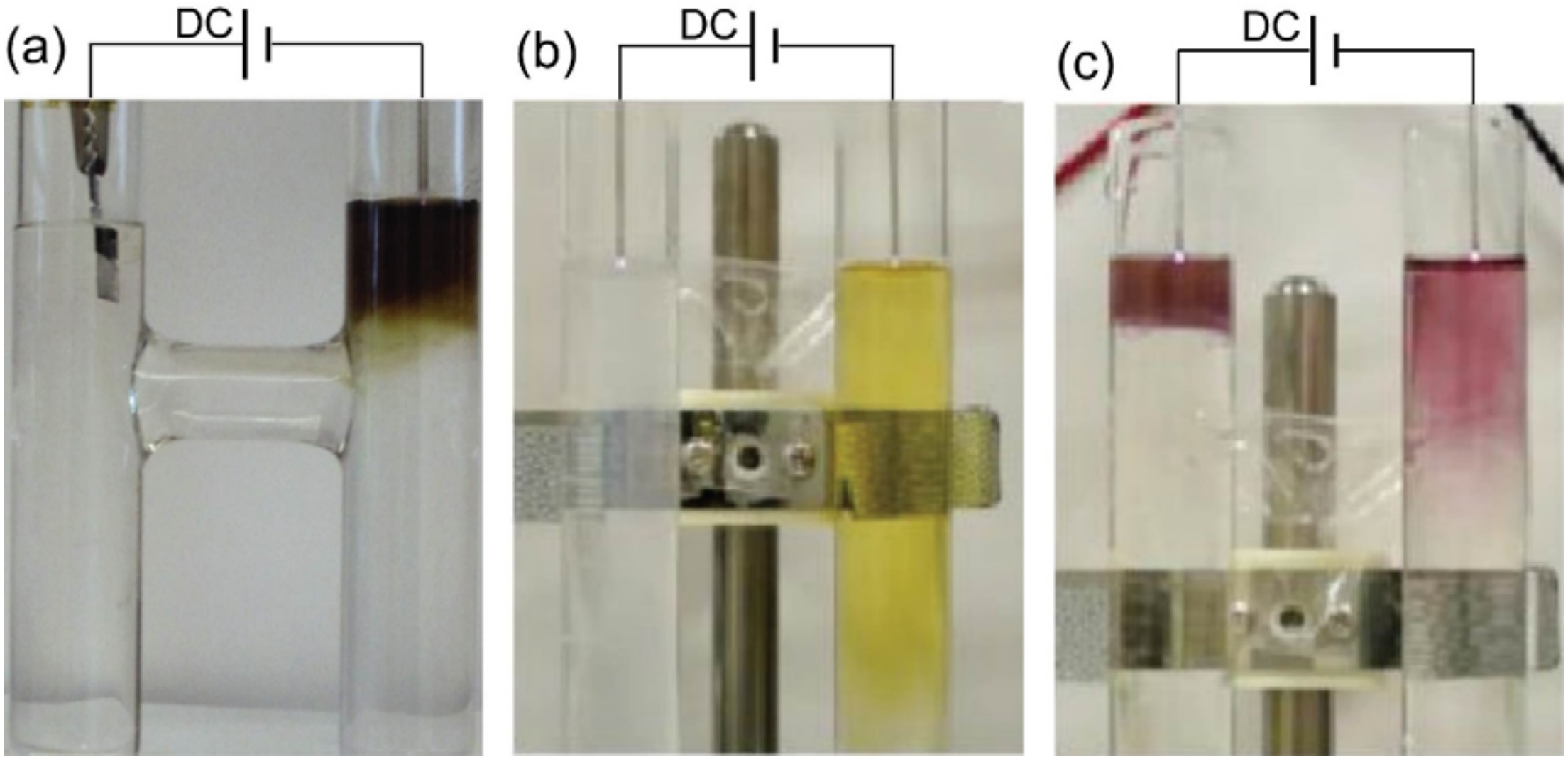

**Figure 8.** Photos of aqueous solutions in H-shape cells. (a) Ag nanoparticles synthesis from Ag foil anodic dissolution induced by cathodic microplasma. The plasma treatment time was 6 min, and fructose was present in solution to prevent agglomeration and precipitation of the particles [69]. (b) Ag and (c) Au nanoparticles synthesized from aqueous solutions of $AgNO_3$ and $HAuCl_4$, respectively, by microplasmas as both anode and cathode simultaneously, and the plasma exposure time was 5 min [154]. Modified with permissions from refs. [69, 154], copyright 2010 IOP Publishing, and copyright from the authors, respectively.

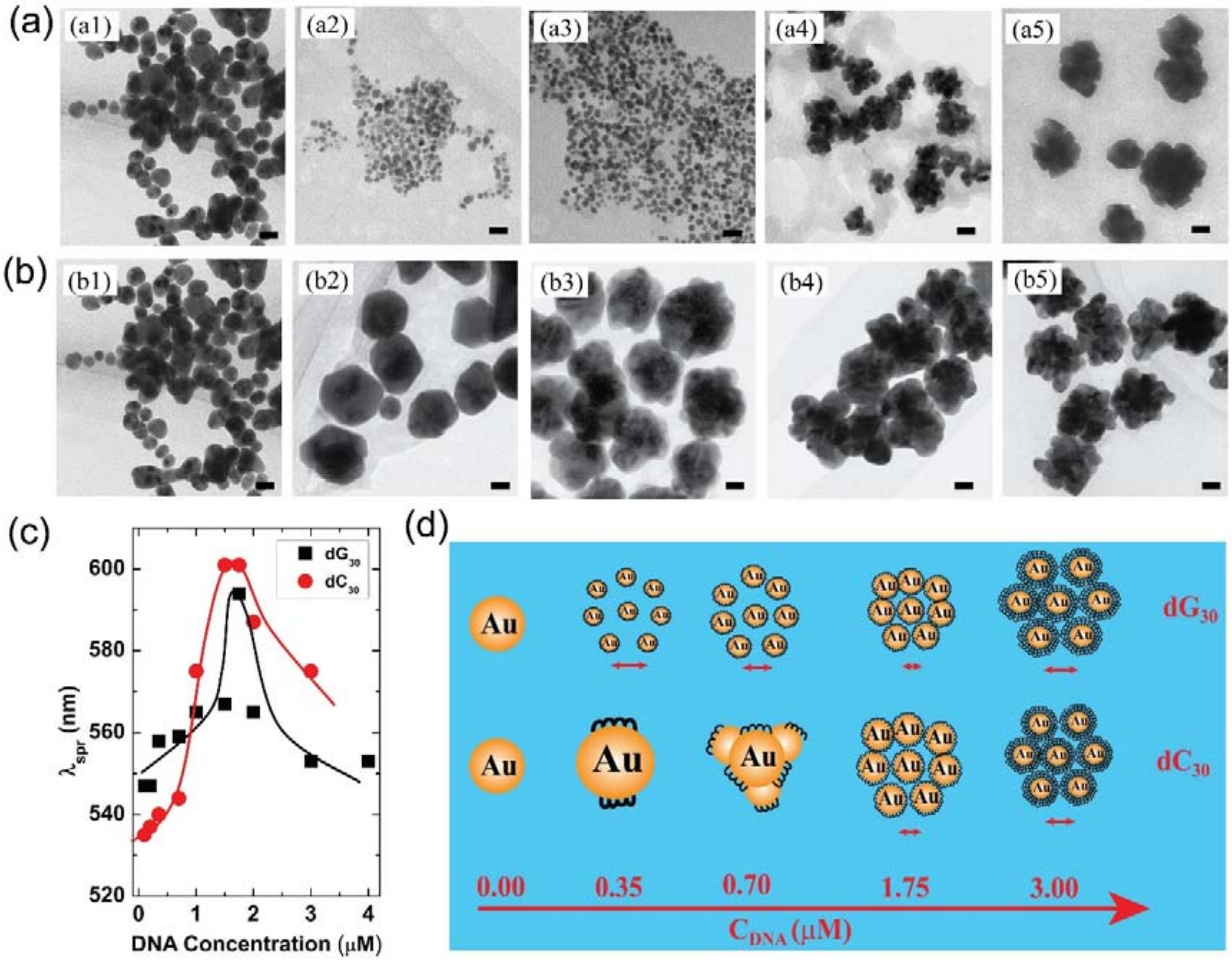

**Figure 9.** TEM images of AuNPs synthesized with 30-mer DNA consisting of (a) guanine ($dG_{30}$) and (b) cytosine ($dC_{30}$) with concentrations of 0, 0.35, 0.70, 1.75, and 3.00 μM for (a1) to (a5) and (b1) to (b5), Scale bar is 20 nm, (c) UV-vis absorption spectra of AuNPs synthesized with different concentration of single-stranded $dG_{30}$ and $dC_{30}$, and (d) schematic of final morphology tuned by DNA concentration and type. Modified with permission from Ref. [71]. Copyright 2012 Elsevier.

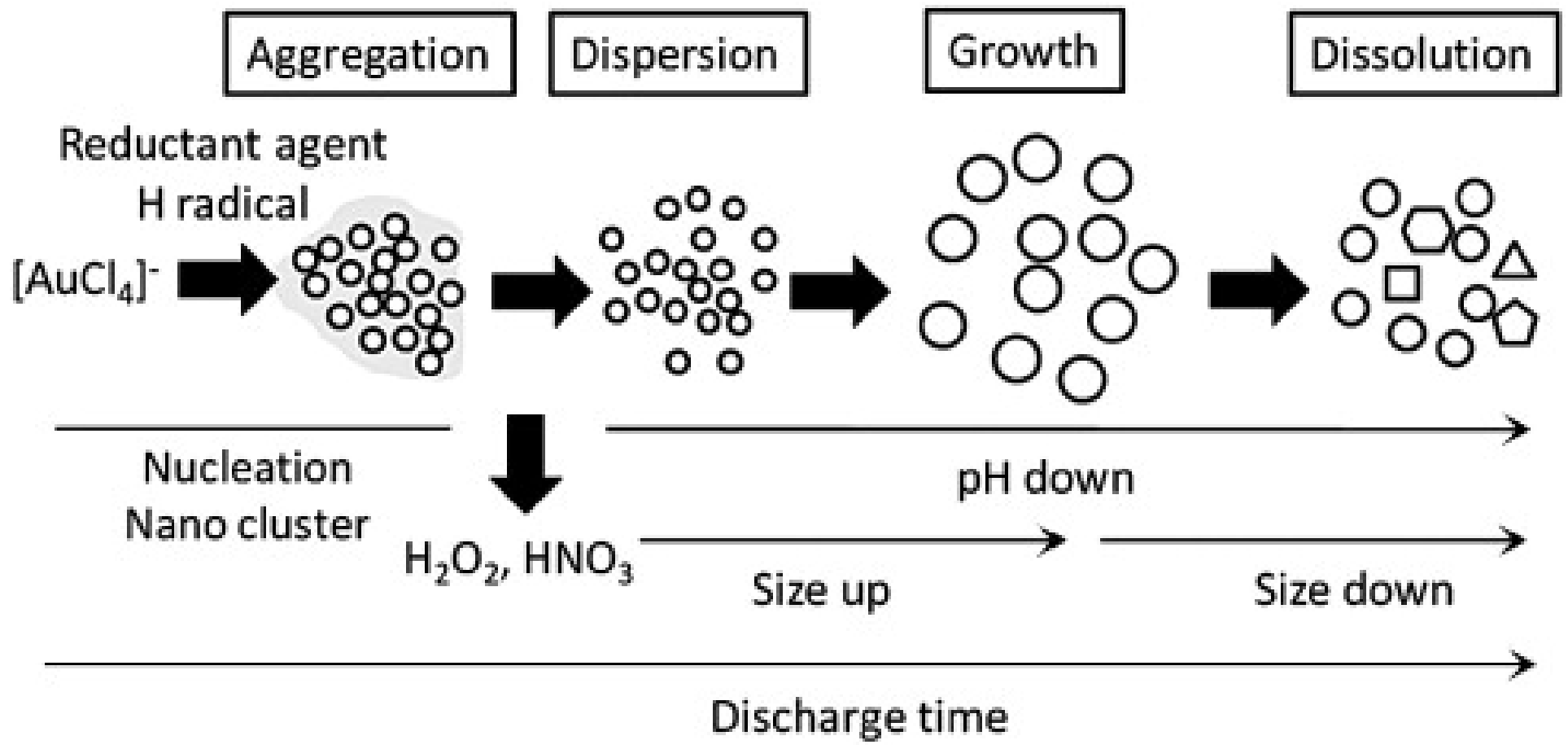

**Figure 10.** Dynamic process of AuNPs synthesis from plasma in liquid. With permission from Ref. [49]. Copyright 2009 Elsevier.

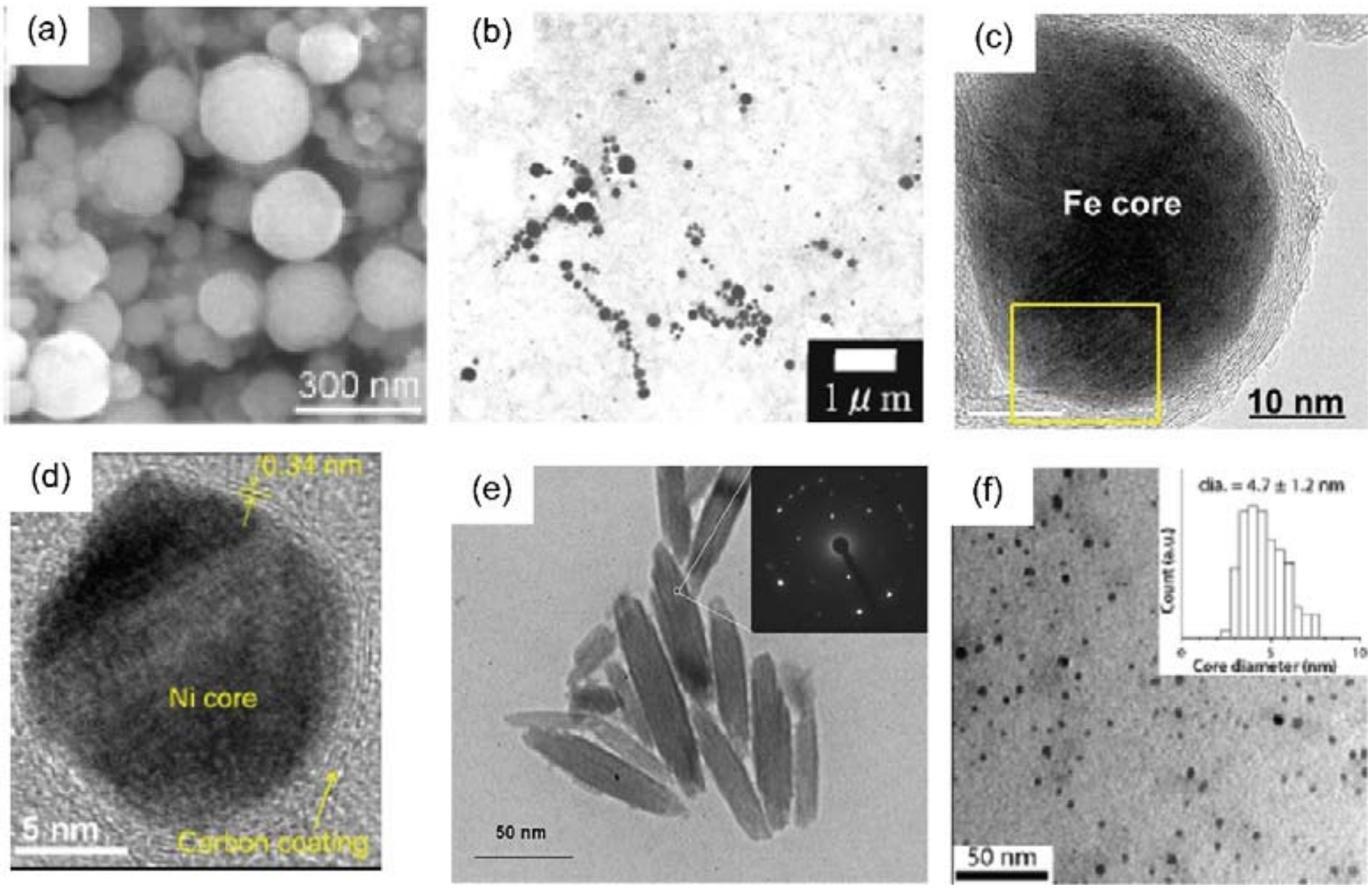

**Figure 11.** (a) SEM and (b) TEM images of NiNPs synthesized from Ni cathode evaporation in $K_2CO_3$ and NaOH solution, respectively, modified with permission from Ref. [123], copyright 2007 American Institute of Physics, and Ref. [125], copyright 2011, Springer. TEM images of (c) Fe@C, and (d) Ni@C NPs synthesized from PLI in ethanol, modified with permissions from Ref. [190], copyright 2011 Elsevier and Ref. [191], copyright 2013 The Japan Society of Applied Physics. (e) TEM image of iron-related NPs synthesized from aqueous solution of $FeCl_3 \cdot 6H_2O$ and polyethylene glycol, modified with permission from Ref. [201], copyright 2011 Elsevier. (f) TEM image of $FeO_x$ NPs synthesized by multi-microplasma-liquid interaction, modified with permission from Ref. [204], copyright from the authors.

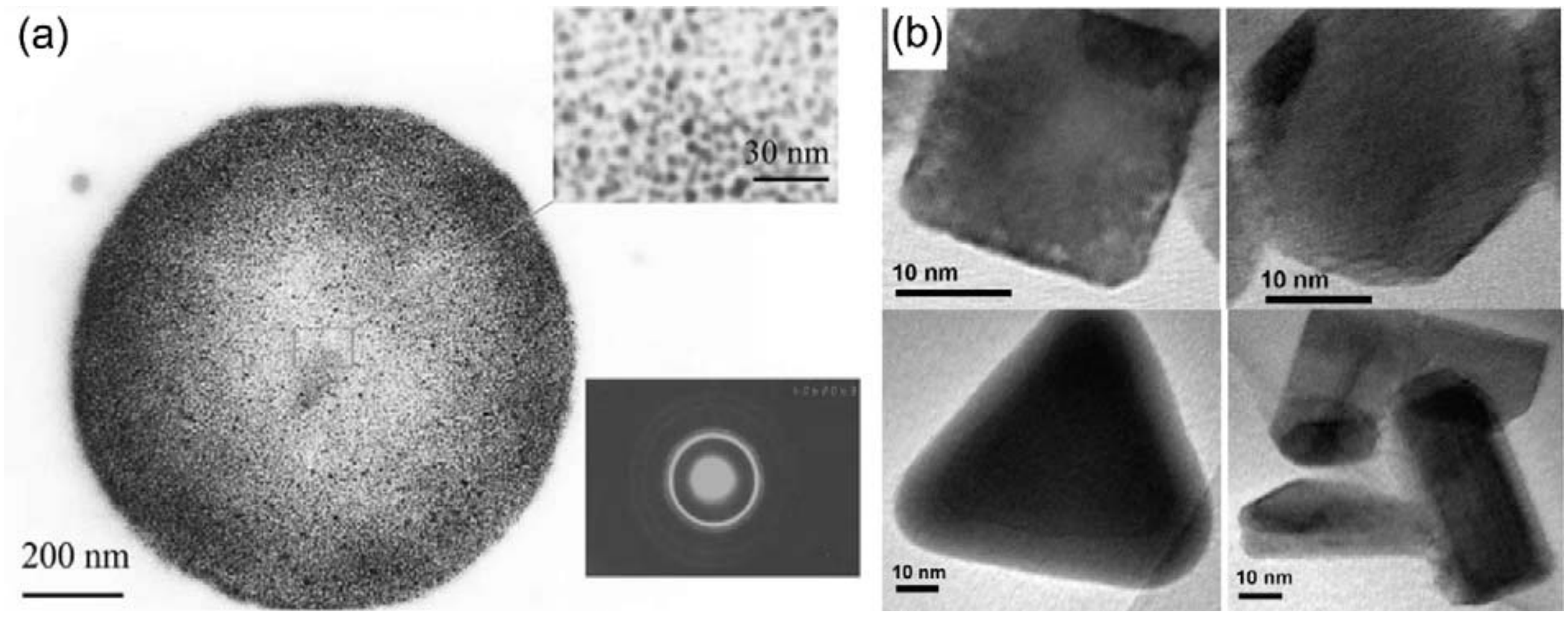

**Figure 12.** (a) TEM image of a self-assembly of CuNPs synthesized from the plasma-liquid interactions in an aqueous solution of ascorbic acid/cetyltrimethylammonium bromide (CTAB) modified with permission from Ref. [214], copyright 2004 Elsevier. (b) TEM images of anisotropic shapes of CuNPs synthesized from aqueous solution of ascorbic acid/gelatin/$CuCl_2$using with 10 min treatment. Ref. [216], copyright 2013 IOP Publishing.

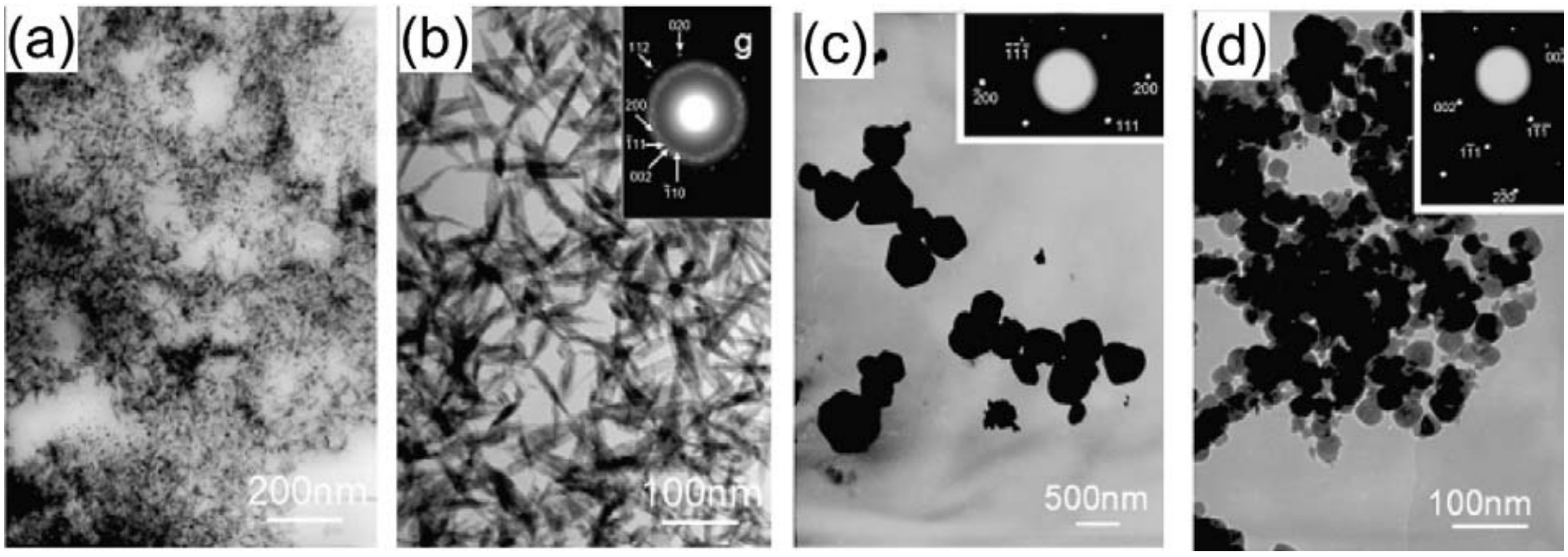

**Figure 13.** TEM images of products from the plasma-liquid interactions (PLIs) after the synthesis of (a) 30 min for CuONPs, (b) 1 week for CuONPs, (c) 2 weeks for $Cu_2$ONPs, and (d) 2 weeks for CuNPs. The PLIs were performed by generating AC plasma over an aqueous solution of $NaNO_3$, Two Cu filaments were used as the electrodes (one over and the other inside the solution). There was no any additive in the solution for the CuONP synthesis, but ascorbic acid or hydrazine hydrate was added to the solution for the $Cu_2$ONP or CuNP synthesis. Modified with permission from Ref. [233], copyright 2005 American Chemical Society.

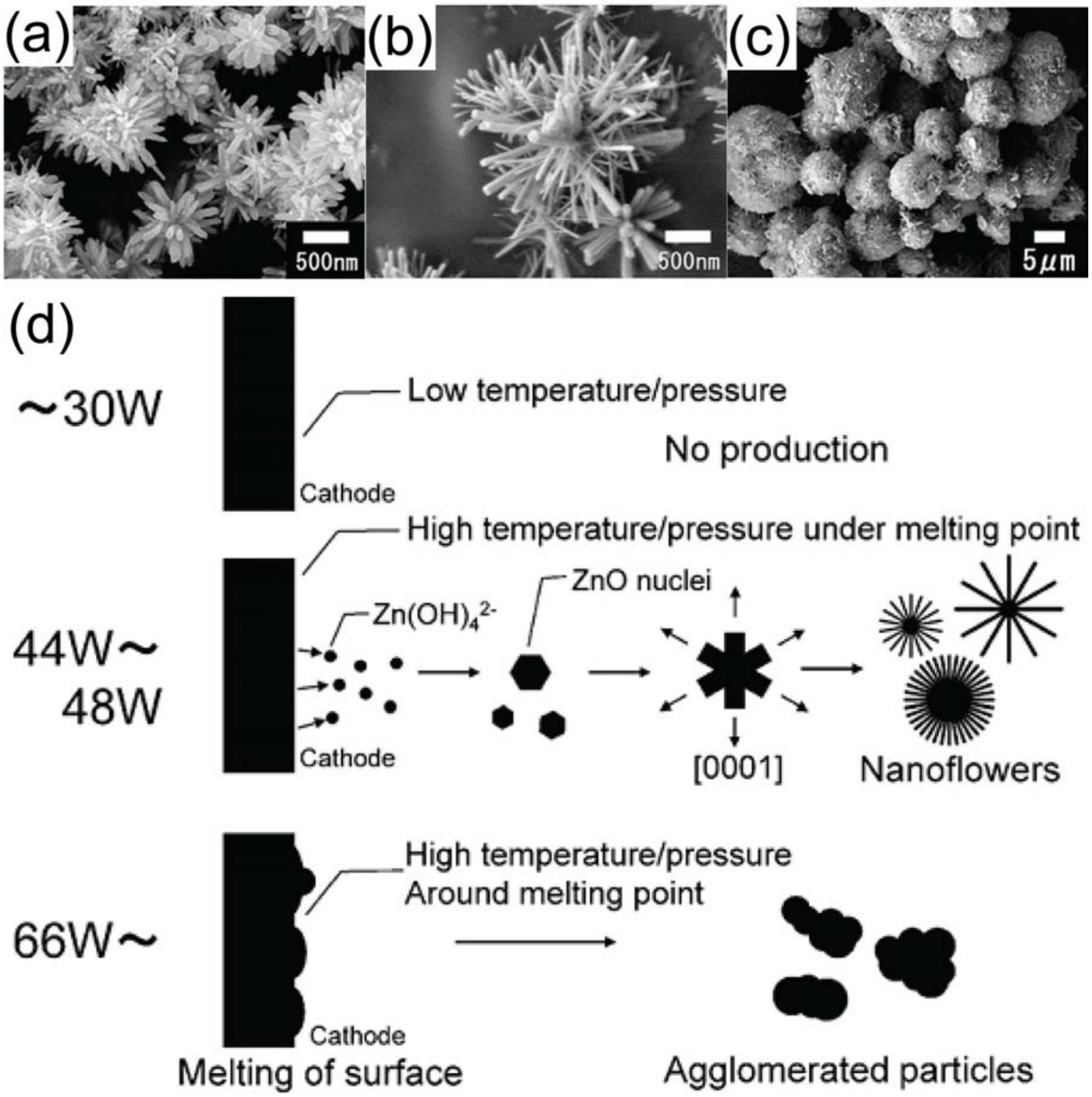

**Figure 14.** SEM images of products obtained from plasma inside liquid, the experimental conditions of $K_2CO_3$ concentration, discharge voltage/power are (a) 1.0 M , 66 V/44 W, (b) 0.5 M , 80 V/48 W, and (c) 0.01 M , 200 V/168 W. (d) Schematic of ZnO synthesized from plasma in liquid. Modified with permission from Ref. [234], copyright 2011 Elsevier.

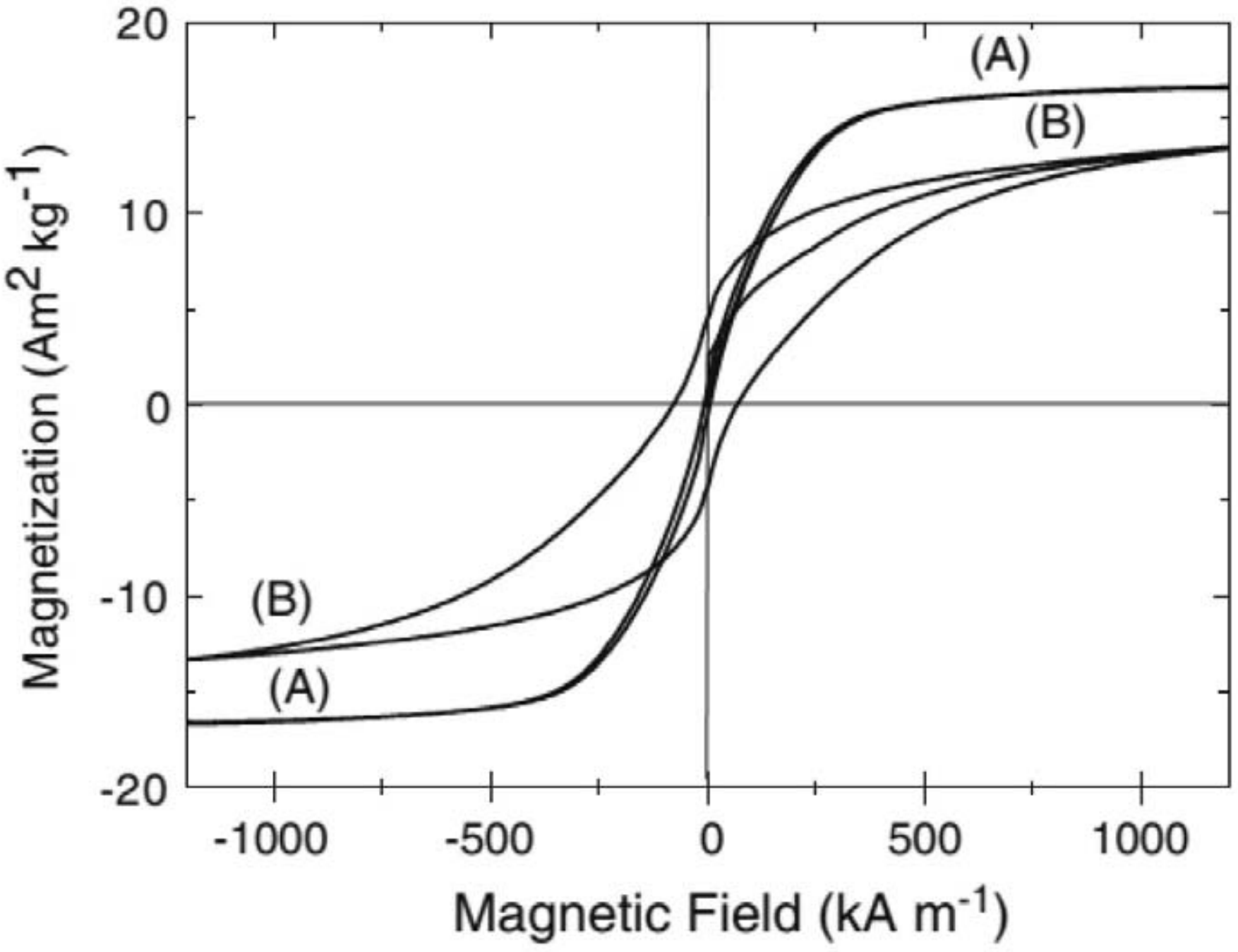

**Figure 15.** Hysteresis loops of the (A) as-synthesized (B) and annealed (at 923 K) Fe-Pt NPs synthesized from the plasma-liquid interactions operated from a $Fe_{50}Pt_{50}$ plate anode and a Fe rod cathode in ethanol. Modified with permission from Ref. [241], copyright 2010 Springer.

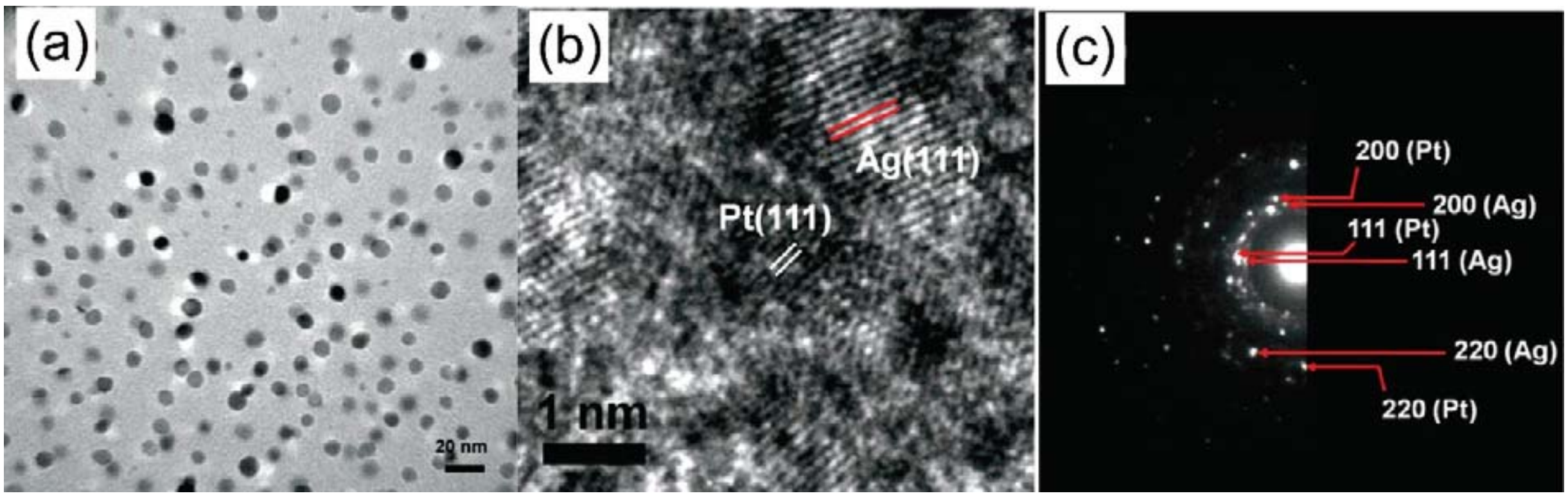

**Figure 16.** (a) TEM image of Ag-Pt NPs, (b) HRTEM image showing many small islands of Pt (dark contrast) on the layer of Ag, and (c) selected area electron diffraction (SAED) patterns of fcc plane reflection of Ag-Pt NPs. The NPs were synthesized from unipolar pulse plasma generated between Ag cathode and Pt anode in an aqueous solution of sodium dodecylsulfonate (SDS, 10 mM) and NaCl (0.1 M) and the discharge time is 30 s. Modified with permission from Ref. [242], copyright 2012 IOP Publishing.

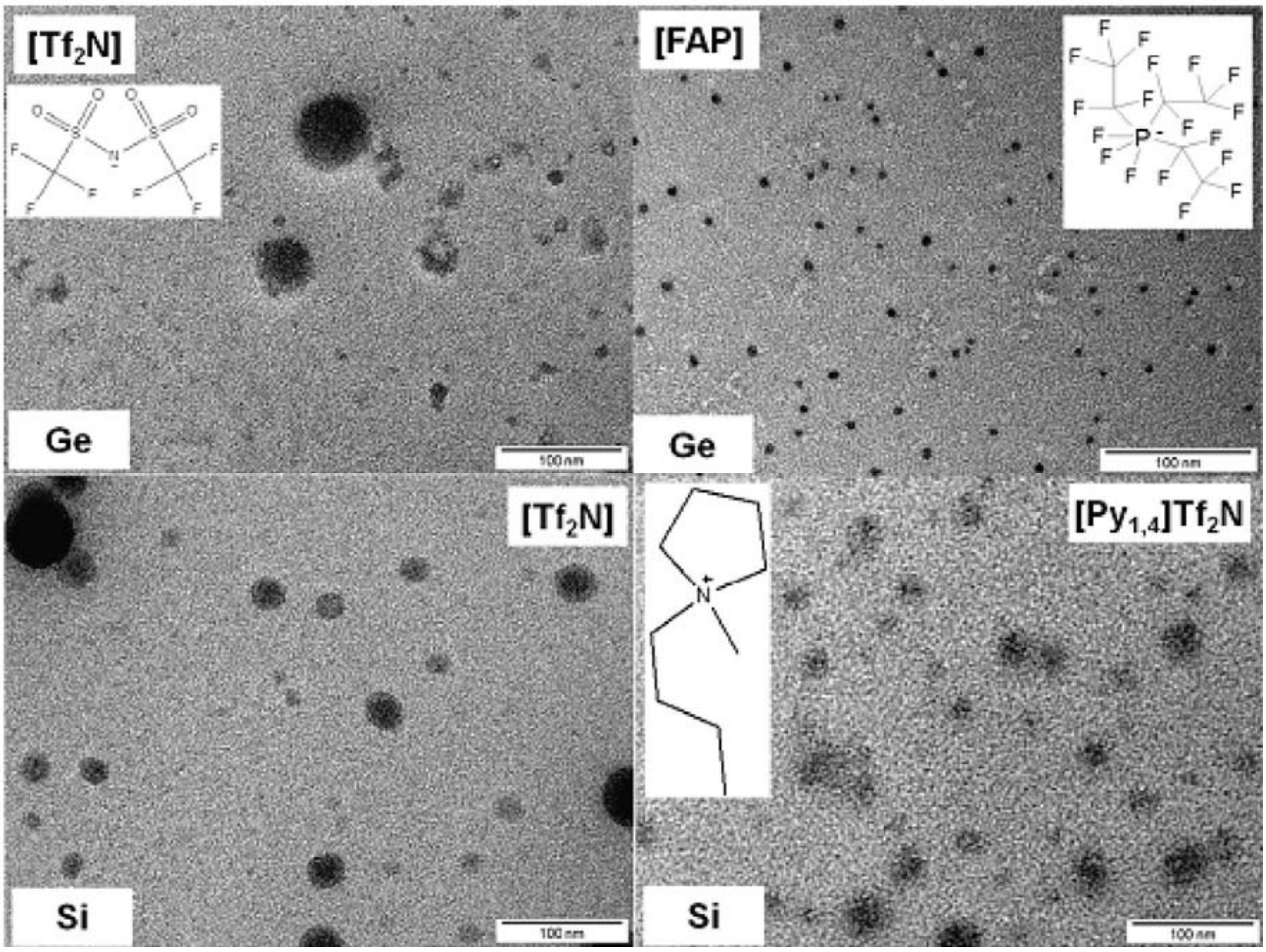

**Figure 17.** TEM images of Si and Ge quantum dots synthesized from a low-pressure plasma-ionic liquid system. $SiCl_4$ and $GeCl_2$ dissolved ionic liquids served as anode. The ionic liquids were 1-ethyl-3-methylimidazolium bis(trifluoromethylsulfonyl)imide ([EMIM]$Tf_2N$), 1-butyl-1-methyl-pyrrolidinium bis(trifluoromethylsulfonyl)amide ([Py1,4]$Tf_2N$), and 1-Ethyl-3-methylimidazolium tris(pentafluoroethyl)trifl uorophosphate ([EMIM]FAP), and the structures of $Tf_2N$, FAP and [Py1,4]$Tf_2N$ are displayed in the image. With permission from Ref. [265], copyright 2013 Elsevier.

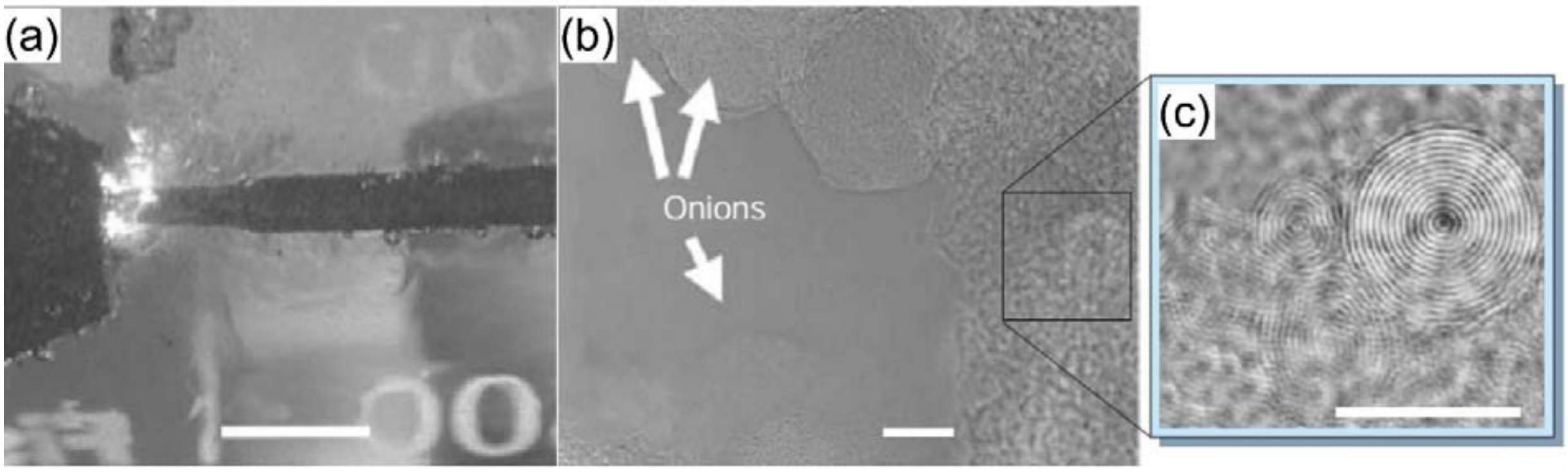

**Figure 18.** (a) A photo of an arc discharge in water for producing carbon onions, scale bar is 12 mm. (b) low- and (c) high-resolution TEM images of formed carbon onions, scale bars is10 nm. Samples were taken from the floating materials on the water surface after their production. With permission from Ref. [52], copyright 2001 Nature Publishing Group.

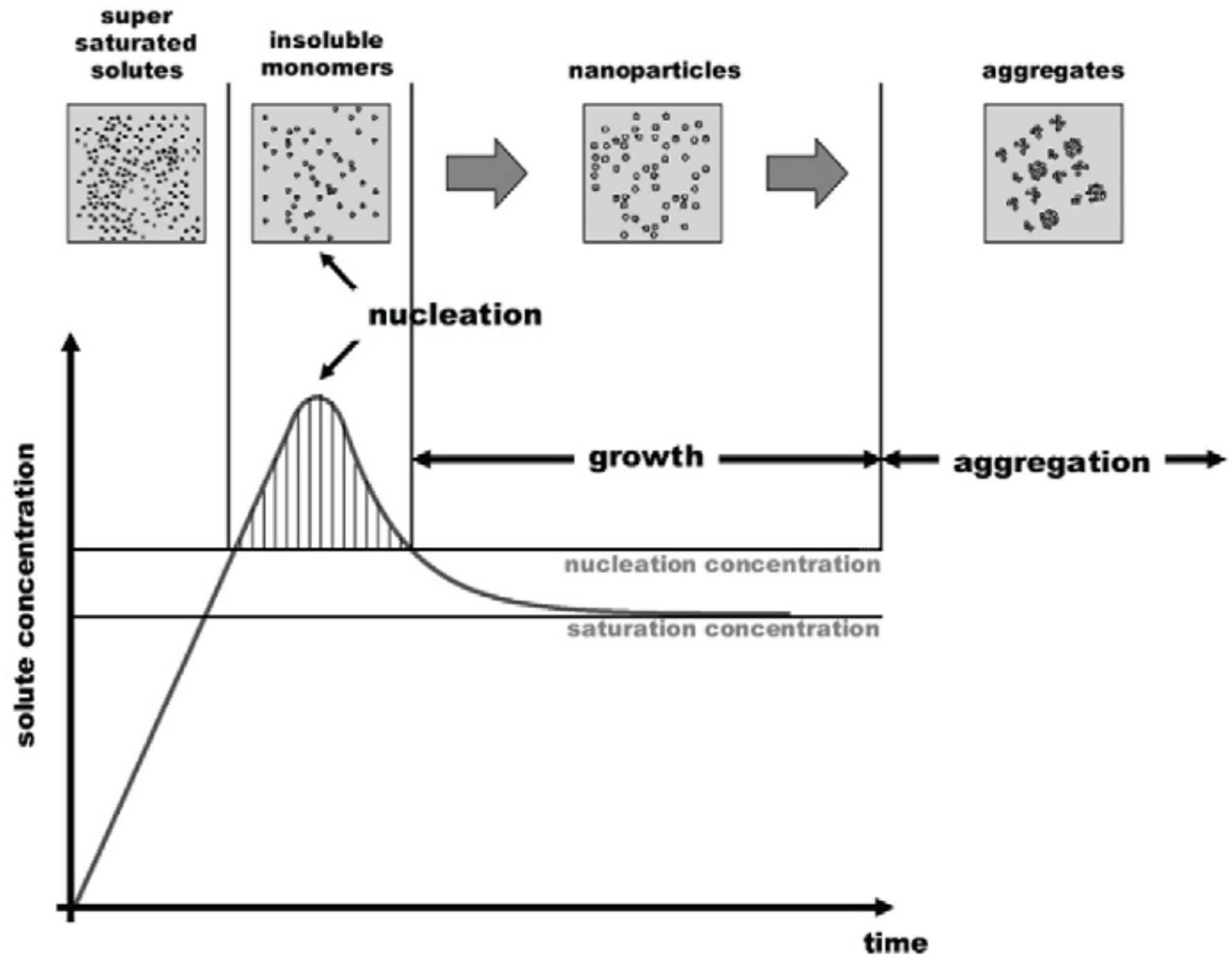

**Figure 19.** Illustration of nucleation and growth for nanoparticle formation in solution. Permitted from Ref. [305], copyright 2004 The Royal Society of Chemistry.